\documentclass[12pt, a4paper, fleqn]{article}

\usepackage[nottoc]{tocbibind} 
\RequirePackage[l2tabu, orthodox]{nag}
\usepackage{silence}
\ActivateWarningFilters[pdftoc]

\usepackage[T1]{fontenc}
\usepackage[utf8]{inputenc}

\usepackage[top=20mm, bottom=20mm, left=25mm, right=25mm]{geometry}

\usepackage{amsmath, amssymb, amsfonts}

\usepackage{pgf}
\usepackage{tikz}
\usepackage{multirow}

\usetikzlibrary{shapes.misc}
\tikzset{cross/.style={cross out, draw=black, thick, fill=none, minimum size=2*(#1-\pgflinewidth), inner sep=0pt, outer sep=0pt}, cross/.default={3pt}}

\usepackage{amsthm}
\theoremstyle{definition}
\newtheorem{theorem}{Claim}

\usepackage[colorlinks=true, linktoc=all, linkcolor=black, citecolor=red, urlcolor=blue, backref=page]{hyperref}

\usepackage{etoolbox}
\makeatletter
\patchcmd{\BR@backref}{\newblock}{\newblock(}{}{}
\patchcmd{\BR@backref}{\par}{)\par}{}{}
\makeatother

\numberwithin{equation}{section}
\renewcommand{\arraystretch}{1.5}

\begin{document}

\

\begin{flushleft}
{\bfseries\sffamily\Large 
Reflecting boundary conditions in critical loop models
\vspace{1.5cm}
\\
\hrule height .6mm
}
\vspace{1.5cm}

{\bfseries\sffamily 
Max Downing,
Jesper Lykke Jacobsen, Rongvoram Nivesvivat, Sylvain Ribault, Hubert Saleur
}
\vspace{3mm}

{\textit{\!\!
(JJ, SR, HS) Institut de physique théorique, CEA, CNRS,
Université Paris-Saclay
\\
(MD, JJ) Laboratoire de Physique de l'École Normale Supérieure, ENS, Université PSL, \\ \hspace{1.65cm}  CNRS, Sorbonne Université, Université Paris Cité, Paris, France
\\
(RN) New York University Abu Dhabi, United Arab Emirates
\\
(HS) Department of Physics and Astronomy, University of Southern California, \\ \hspace{.8cm} Los Angeles
}}
\vspace{2mm}

{\textit{E-mail:} \texttt{max.downing@phys.ens.fr,
jesper.jacobsen@ens.fr, rongvoramnivesvivat@gmail.com, sylvain.ribault@ipht.fr, hubert.saleur@ipht.fr
}}
\end{flushleft}
\vspace{7mm}

{\noindent\textsc{Abstract:}
In critical loop models, we call a boundary sticky if loops can attach to it, and reflecting otherwise.
Using analytic bootstrap methods, we show that reflecting boundaries are characterised by one complex parameter, analogous to the boundary cosmological constant in Liouville theory. We determine disc 1-point functions, and write an explicit formula for disc 2-point functions as infinite combinations of conformal blocks. We also sketch the lattice interpretation of reflecting boundaries.
}

\clearpage

\hrule 
\tableofcontents
\vspace{5mm}
\hrule
\vspace{5mm}

\hypersetup{linkcolor=blue}

\section{Introduction}

In two-dimensional statistical physics, ensembles of non-intersecting loops describe a wide range of phenomena, including percolation and the physics of polymers. At the lattice level, these models are accessible to powerful algebraic and integrable techniques; see \cite{jac09} for a review. At criticality, loop models give rise to conformal field theories (CFTs), and exact results can be obtained using algebraic CFT methods \cite{jac09,car05}, probabilistic descriptions in terms of Schramm--Loewner evolutions \cite{car05,law07}, or Conformal Loop Ensembles \cite{she06}.

More recently, analytic bootstrap methods have led to substantial progress in the computation of correlation functions in critical loop models \cite{rib24}. In particular, analytic formulas have been obtained for a large class of bulk structure constants \cite{nrj23,jnrr25}. Moreover, we now understand that several apparently independent models, including the $O(n)$, Potts and $PSU(n)$ models, can be treated in the same way, and even share some correlation functions.
Nevertheless, critical loop models remain far from being fully solved. Even basic structural properties, such as the existence of operator product expansions, remain mysterious.

\subsubsection{Boundary critical loop models}

In this work, we begin applying the same bootstrap methods to critical loop models on surfaces with boundaries. Boundary loop models are expected to contain a rich variety of boundary conditions and boundary fields, and their correlation functions describe geometrical observables that do not arise on the sphere.

Some well-known exact results already illustrate the importance of boundary correlation functions. These include:
\begin{itemize}
\item Cardy's computation of the probability that a cluster in critical percolation connects two disjoint boundary segments. This calculation is based on the determination of a boundary four-point function \cite{car91}.
\item Schramm's computation of the probability that a percolation hull running between two distinct boundary points passes to the left of a given bulk point. This calculation is based on a correlation function involving two boundary fields and one bulk field \cite{schramm01}.
\end{itemize}
The results of Cardy and Schramm relied on formulating geometrical problems in terms of correlation functions containing degenerate fields. The corresponding correlators obey simple differential equations, and the resulting probabilities are expressed in terms of hypergeometric functions.

Boundary critical loop models contain many more fields besides the degenerate fields, leading to a much larger class of correlation functions and geometrical observables. In particular, in this article we determine a family of bulk two-point functions on the disc, given in \eqref{2pt-bulk}. Unlike Cardy's and Schramm's probabilities, our disc two-point functions are infinite linear combinations of Virasoro conformal blocks, and do not obey any known differential equation. Overcoming the technical limitations of earlier works therefore gives us access to more general observables. These include the probability that a critical percolation cluster connects two bulk points on the disc, which is studied more specifically in a separate text \cite{djns26}.

\subsubsection{Terminology}

Two-dimensional statistical models of loops are theories of non-crossing random curves that can be formulated either on a lattice or in the continuum limit.
These curves are fractals in the critical case. Although their local properties do not depend on global, topological considerations, we will need to distinguish
topologies when discussing correlation functions and boundary
conditions. When this distinction is needed, we shall use the following terminology:
\begin{itemize}
 \item A {\bf loop} is a closed curve (i.e.\ it has the topology of a circle).
 \item An {\bf arc} is an open curve with two end points (i.e.\ it has the topology of an interval).
\end{itemize}
So loop models describe random curves which can be topologically either loops or arcs.
Notice that what we here call an {\em arc} was previously named `open loop segment' by some of the authors \cite{jnrr25}.

\subsubsection{Reflecting and sticky boundaries}

In a statistical model of loops, the behaviour of a curve near a boundary depends on whether the curve is attached to the boundary, or may  move freely. We take this as a fundamental property of the boundary, and call the boundary sticky or reflecting depending on this property. This property can be reformulated in terms of the degrees of freedom of a boundary point that is visited by a curve:
\begin{itemize}
 \item $2$ degrees of freedom if the curve can detach,
 \item $1$ if the curve can slide along the boundary, but must remain attached to it,
 \item $0$ if the curve is attached to a fixed boundary point, which is then interpreted as the insertion point of a boundary field.
\end{itemize}
\begin{align}
\renewcommand{\arraystretch}{1.4}
 \begin{array}{|c||c|c|c|}
 \hline
  \# \text{d.o.f.} & 2 & 1 & 0
  \\
  \hline
  \text{Name} & \text{Reflecting boundary} & \text{Sticky boundary} & \text{Boundary field}
  \\
  \hline
   \text{Diagram} &
   \begin{tikzpicture}[baseline = (base), scale = .3]
     \coordinate (base) at (0, 1.5);
     \node at (0, 4.3) {};
     \clip (-5, -1) rectangle (5, 4);
     \draw(-5, 0) -- (5, 0);
     \draw[thick, red] plot [smooth] coordinates {(-1, 0)(-.7, 2)(-2.7, 4)};
     \draw[thick, red] plot [smooth] coordinates {(-1, 0)(-1.7, 2)(-4.5, 4)};
     \draw[red!60] plot [smooth] coordinates {(-3.3, 4)(-1, 2)(-1.4, .3)(-2.2, 2)(-4.9, 4)};
     \draw[red!60] plot [smooth] coordinates {(-3.7, 4)(-1.5, 2.2)(-2, .5)(-3, 2.1)(-6, 4)};
     \draw[thick, red] plot [smooth] coordinates {(0, 4)(1.2, 2)(1, 1)(2, 2)(4.5, 4)};
     \draw[red!60] plot [smooth] coordinates {(0.4, 4)(1.6, 2)(1.4, .8)(2.5, 2)(5, 4)};
     \draw[red!60] plot [smooth] coordinates {(0.8, 4)(2, 2)(1.8, .6)(3, 2)(6, 4)};
   \end{tikzpicture}
   &
   \begin{tikzpicture}[baseline = (base), scale = .3]
     \coordinate (base) at (0, 1.5);
     \node at (0, 4.3) {};
     \clip (-5, -1) rectangle (5, 4);
     \draw(-5, 0) -- (5, 0);
     \draw[thick, red] plot [smooth] coordinates {(-1, 0)(-.7, 2)(-2.7, 4)};
     \draw[thick, red] plot [smooth] coordinates {(-1, 0)(-1.7, 2)(-4.5, 4)};
     \draw[red!60] plot [smooth] coordinates {(-3.3, 4)(-1, 2)(-1.6, 0)(-2.2, 2)(-4.9, 4)};
     \draw[red!60] plot [smooth] coordinates {(-3.7, 4)(-1.5, 2.2)(-2.2, 0)(-3, 2.1)(-6, 4)};
     \draw[thick, red] plot [smooth] coordinates {(1, 4)(2.4, 2)(2, 0)};
     \draw[red!60] plot [smooth] coordinates {(.7, 4)(2, 2)(1.4, 0)};
     \draw[red!60] plot [smooth] coordinates {(.4, 4)(1.6, 2)(.8, 0)};
     \begin{scope}[shift = {(2.5, .4)}]
     \draw[thick, red] plot [smooth] coordinates {(0, 4)(1.2, 2)(1, 1)(2, 2)(4.5, 4)};
     \draw[red!60] plot [smooth] coordinates {(0.4, 4)(1.6, 2)(1.4, .8)(2.5, 2)(5, 4)};
     \draw[red!60] plot [smooth] coordinates {(0.8, 4)(2, 2)(1.8, .6)(3, 2)(6, 4)};
     \end{scope}
     \clip (-5, 0) rectangle (5, -1);
     \node[fill, circle, minimum size = 1.4mm, inner sep = 0] at (-1, 0) {};
     \node[fill=black!40, circle, minimum size = 1.4mm, inner sep = 0] at (-1.6, 0) {};
     \node[fill=black!40, circle, minimum size = 1.4mm, inner sep = 0] at (-2.2, 0) {};
      \node[fill, circle, minimum size = 1.4mm, inner sep = 0] at (2, 0) {};
     \node[fill=black!40, circle, minimum size = 1.4mm, inner sep = 0] at (1.4, 0) {};
     \node[fill=black!40, circle, minimum size = 1.4mm, inner sep = 0] at (.8, 0) {};
   \end{tikzpicture}
   &
   \begin{tikzpicture}[baseline = (base), scale = .3]
     \coordinate (base) at (0, 1.5);
     \node at (0, 4.3) {};
     \clip (-5, -1) rectangle (5, 4);
     \draw(-5, 0) -- (5, 0);
     \draw[thick, red] plot [smooth] coordinates {(0, 0)(2, 2)(1.8, 4)};
     \draw[red!60] plot [smooth] coordinates{(0, 0)(1.6, 2)(1.2, 4)};
     \draw[red!60] plot [smooth] coordinates{(0, 0)(1.2, 2)(1, 4)};
     \draw[thick, red] plot [smooth] coordinates {(0, 0)(.5, 2)(-2, 4)};
     \draw[red!60] plot [smooth] coordinates {(0, 0)(.1, 2)(-2.5, 4)};
     \draw[red!60] plot [smooth] coordinates {(0, 0)(-.3, 2)(-3, 4)};
     \draw[thick, red] plot [smooth] coordinates {(0, 0)(-4, 2)(-5.5, 4)};
     \draw[red!60] plot [smooth] coordinates {(0, 0)(-3.5, 2)(-5, 4)};
     \draw[red!60] plot [smooth] coordinates {(0, 0)(-3, 2)(-4, 4)};
     \begin{scope}[shift = {(2.5, .4)}]
     \draw[thick, red] plot [smooth] coordinates {(0, 4)(1.2, 2)(1, 1)(2, 2)(4.5, 4)};
     \draw[red!60] plot [smooth] coordinates {(0.4, 4)(1.6, 2)(1.4, .8)(2.5, 2)(5, 4)};
     \draw[red!60] plot [smooth] coordinates {(0.8, 4)(2, 2)(1.8, .6)(3, 2)(6, 4)};
     \end{scope}
     \node[fill, circle, minimum size = 1.4mm, inner sep = 0] at (0, 0) {};
   \end{tikzpicture}
   \\
   \hline
 \end{array}
\end{align}
An arc can end at a boundary field, or on a sticky boundary, but not on a reflecting boundary.
A loop can touch any boundary in lattice loop models, but in some cases it will shy away from it in the continuum limit.

In two-dimensional CFT, boundary conditions are characterised by the disc 1-point functions of bulk fields. In critical loop models, we distinguish two types of bulk fields, called diagonal and non-diagonal. A non-diagonal field $V_{(r,s)}$ corresponds to a puncture with $2r\in\mathbb{N}^*$ legs, i.e. a puncture that emits $2r$ arcs. A diagonal field $V_P$ with $P\in\mathbb{C}$ changes the weight of any loop that surrounds it to a value $w(P)$. (See Table \ref{tab:bulk fields} for details.)
\begin{align}
\renewcommand{\arraycolsep}{2cm}
\begin{array}{cc}
 \begin{tikzpicture}[baseline=(current bounding box.center), scale=1.7]
% smoother, wider-than-tall loop around the point
\draw[thick, red] plot[smooth cycle, tension=.85] coordinates {
  (0, .45)
  (.42, .34)
  (.62, .08)
  (.50, -.24)
  (.12, -.42)
  (-.32, -.36)
  (-.70, -.08)
  (-.46, .26)
};
\node at (.38, .56) {$w(P)$};
% central point and label
\draw (0, 0) node [cross]{};
\node[left] at (0, 0) {$V_P$};
\end{tikzpicture}
&
\begin{tikzpicture}[baseline=(current  bounding  box.center), scale = 1.2]
\draw[thick, red] plot [smooth] coordinates {(0, 0)(.2, .7)(.4, 1.2)};
\node at (.5, 1.3) {$\scriptstyle{1}$};
\draw[thick, red] plot [smooth] coordinates {(0, 0)(.3, .6)(.6, 1)};
\node at (.7, 1.1) {$\scriptstyle{2}$};
\draw[thick, red] plot [smooth] coordinates {(0, 0)(.4, .5)(.8, .8)};
\node at (.9, .9) {$\scriptstyle{3}$};
\draw[thick, red] plot [smooth] coordinates {(0, 0)(.5, -.3)(1.1, -.5)};
\node at (1.3, -.5) {$\scriptstyle{2r}$};
 \draw (0, 0) node[fill, circle, minimum size = 1.6mm, inner sep = 0]{};
 \node[left] at (0, 0) {$V_{(r,s)}$};
 \draw[dashed] plot [smooth] coordinates {(.6, .45)(.75, .1)(.7, -.25)};
\end{tikzpicture}
\\
\text{Diagonal field} & \text{Non-diagonal field}
\end{array}
\end{align}
Disc 1-point functions of non-diagonal fields vanish if the boundary is reflecting,
\begin{align}
\left<V_{(r,s)}\right>_{\text{reflecting}}=0\ ,
\end{align}
and this condition characterises reflecting boundaries in the conformal bootstrap approach.
Let us illustrate some examples of disc 1-point and 2-point functions in the presence of reflecting and sticky boundaries:
\begin{align}
\begin{array}{c}
\begin{array}{cc}
\begin{tikzpicture}[baseline=(base), scale = .5]
    \coordinate (base) at (0, 0);
  \draw (1.5, 0) node [cross]{};
    \draw (1.5, 0) ellipse (2.6cm and 2.6cm);
  \end{tikzpicture}
  &
\begin{tikzpicture}[baseline=(base), scale = .5]
    \coordinate (base) at (0, 0);
  \node[fill, circle, minimum size = 1.6mm, inner sep = 0] at (0, 0) (a) {};
  \node[fill, circle, minimum size = 1.6mm, inner sep = 0] at (3, 0) (c) {};
    \draw[thick, red] (a) -- (c);
    \draw[thick, red] (a) to [out = -15, in = -90] (3.5, 0) to [out = 90, in = 15] (a);
    \draw (1.5, 0) ellipse (2.6cm and 2.6cm);
  \end{tikzpicture}
\end{array}
\\
\text{Reflecting boundary}
\end{array}
\hspace{1cm}
\begin{array}{c}
\begin{array}{cc}
  \begin{tikzpicture}[baseline=(current  bounding  box.center), scale = .5]
%\draw[thick, red] plot [smooth] coordinates {(0, 0)(.5, 1.2)(1.3, 2.5)};
\draw[thick, red] plot [smooth] coordinates {(0, 0)(1.4, 1.5)(2.1, 2.)};
\draw[thick, red] plot [smooth] coordinates {(0, 0)(1.4, .8)(2.1, 2.)};
\draw[thick, red] plot [smooth] coordinates {(0, 0)(1.1, -1.3)(2., -2.)};
 \draw (0, 0) node[fill, circle, minimum size = 1.6mm, inner sep = 0]{};
 \draw (0.4, 0.1) ellipse (2.6cm and 2.6cm);
\end{tikzpicture}
&
\begin{tikzpicture}[baseline=(base), scale = .5]
    \coordinate (base) at (0, 0);
  \draw (3, 0) node [cross]{};
    \draw[thick, red] (0, 0) to [out = -15, in = -90] (3.5, 0) to [out = 90, in = 15] (0,0);
    \draw[thick, red] plot [smooth] coordinates {(0, 0)(.5, 1.2)(1.7, 2.5)};
    \draw[thick, red] plot [smooth] coordinates {(0, 0)(1, -.7)(3.5, -1.6)};
    \node[fill, circle, minimum size = 1.6mm, inner sep = 0] at (0, 0) {};
    \draw (1.5, 0) ellipse (2.6cm and 2.6cm);
  \end{tikzpicture}
\end{array}
\\
\text{Sticky boundary}
\end{array}
\end{align}
In this work, we focus on reflecting boundary conditions, which are technically simpler than sticky boundary conditions. Reflecting boundaries already include physically important examples, such as free and wired boundaries in the critical Potts model \cite{djns26}.

Except in Section \ref{sec:latt}, the technical results of this work rely entirely on the bootstrap method. We regard critical loop models as CFTs characterised by their fields and their correlation functions. Nevertheless, we use intuition from lattice loop models, in particular when it comes to the fields' legs.

\subsubsection{Outline of the paper}

In section \ref{sec:main results}, we present our main results:
\begin{itemize}
    \item The derivation of two classes of boundary conditions, reflecting and sticky.
    \item The disc 1-point functions of bulk fields for reflecting boundary conditions.
    \item  An explicit expression \eqref{2pt-bulk} for the disc 2-point function of bulk fields, as a linear combination of bulk channel conformal blocks.
    \item The boundary channel spectrum \eqref{s12} for the disc 2-point functions, which allows us to check crossing symmetry.
\end{itemize}
In section \ref{sec:analytic bootstrap}, we derive these results analytically. We review the bulk structure constants that enter the disc 2-point functions. We then derive the distinction between reflecting and sticky boundary conditions. Next we derive shift equations for the disc 1-point functions of diagonal fields and solve them. Then we compute the annulus partition function and use its modular transformation to determine the spectrum of boundary fields. Finally we use bulk-to-boundary fusion rules to determine which subset of this spectrum contributes to a given disc 2-point function.

In section \ref{sec:num checks} we test the analytic results using the numerical bootstrap. We briefly explain how crossing symmetry gives rise to a finite system of linear equations. We then provide numerical evidence that the proposed bulk-channel decomposition is compatible with the predicted boundary spectrum.

In section \ref{sec:latt}, we construct reflecting boundary conditions in lattice loop models. We also classify the already known boundary conditions as reflecting or sticky.

In this article, we assume some familiarity with 2-dimensional boundary CFT and with critical loop models on the sphere. For boundary CFT, we recommend the classic work of Lewellen \cite{lew92}. For critical loop models on the sphere, we recommend the review \cite{rib24}.

\section{Main results}\label{sec:main results}

We begin with our notations, and a brief discussion of bulk field normalisations.

\subsubsection{Notations}

We use the following notations for the central charge $c$ and conformal dimension $\Delta$:
\begin{align}
 c= 13-6\beta^2-6\beta^{-2} \quad ,\quad \Delta = \frac{c-1}{24}+P^2 \quad ,\quad P_{(r,s)} = \frac12\left(\beta r-\beta^{-1}s\right)\ ,
 \label{cdp}
\end{align}
where $P$ is the momentum and $r,s$ are Kac indices. The central charge is a complex number obeying $\Re c<13\iff \Re\beta^2>0$. This is necessary for the convergence of OPEs \cite{rib24}. Note that $\mathbb{N} = \{0,1,2,\dots\}$ and $\mathbb{N}^* = \{1,2,\dots\}$.

In critical loop models, the bulk theory has three types of primary fields. They are characterised by their left and right conformal dimensions $(\Delta,\bar\Delta)$, whose difference is the conformal spin $S = \Delta - \bar\Delta$. We list the fields below and give their left and right momenta:
\begin{align}\label{tab:bulk fields}
\renewcommand{\arraystretch}{1.5}
 \begin{array}{|r|l|l|l|l|}
  \hline
  \text{Name} & \text{Notation} & \text{Conditions} & (P,\bar P) & \text{Spin}
  \\
  \hline \hline
  \text{Degenerate} &  V^d_{\langle 1,s\rangle} &  s\in\mathbb{N}^* & \left(P_{(1,s)},P_{(1,s)}\right) & 0
  \\
  \hline
  \text{Diagonal} & V_P & P\in\mathbb{C} & \left(P, P\right) & 0
  \\
  \hline
  \text{Non-diagonal} & V_{(r,s)} &
   r\in\frac12\mathbb{N}^*, \ s\in\frac{1}{r}\mathbb{Z} & \left(P_{(r,s)},P_{(r,-s)}\right) & -rs
  \\
  \hline
 \end{array}
\end{align}
Diagonal fields $V_P$ change the weight of a loop that surrounds them from $n$ to $w(P)$, where
\begin{align}
 n = 2\cos(2\pi P_{(1,1)} \beta)=-2\cos(\pi\beta^2) \quad , \quad w(P) = 2\cos(2\pi P \beta)\ .
 \label{nwp}
\end{align}

\subsubsection{Bulk field normalisations}

For non-diagonal fields, we will be using the same normalisation as was used previously in \cite{nrj23,rib24}. However, for diagonal fields, we will use a different normalisation.
If $\tilde V_P$ is a diagonal field from \cite{nrj23,rib24}, then it is related to our diagonal field $V_P$ by the rescaling
\begin{equation}\label{eq:diag rescale}
    V_P = \lambda_P \tilde V_P \;,\quad \lambda_P = \frac{128 P}{\beta - \beta^{-1}} \prod_\pm \frac{\sin(2\pi P \beta^{\pm1}) \Gamma_\beta(\pm2P) \Gamma_\beta(\beta^{\pm1})^2}{\sin(\pi \beta^{\pm2})\Gamma_\beta(2\beta^{\pm1})} \;.
\end{equation}
In \cite{nrj23,rib24} diagonal fields are even in their momentum, $\tilde V_P = \tilde V_{-P}$. Since $\lambda_P$ is odd, the diagonal fields in this article are odd in their momentum
\begin{equation}\label{odd}
    V_P = -V_{-P} \;.
\end{equation}
In \cite{nrj23,rib24} the normalisation of diagonal and non-diagonal fields was such that if a quantity was given for a non-diagonal field $V_{(r,s)}$, then the analogous expression could be found for a diagonal field $\tilde V_P$ using the substitution $(r,s) = (0,2P\beta)$. Schematically this is equivalent to the identity $V_{(0,2P\beta)} = \tilde V_P$. Using \eqref{eq:diag rescale} this relationship becomes
\begin{equation}\label{eq:non-diag to diag}
    V_{(0,2P\beta)} = \lambda_P^{-1} V_P \;.
\end{equation}

\subsection{Boundary conditions and disc 1-point functions}

In section \ref{sec:dand}, we will
constrain boundary conditions by analysing the disc 1-point functions of bulk fields $\left<V\right>_\sigma$, where $\sigma $ is a formal parameter for boundary conditions.
By conformal symmetry, the 1-point function vanishes for all fields with nonzero spins: in our spectrum \eqref{tab:bulk fields}, this means $rs\neq 0\implies \left<V_{(r,s)}\right>_\sigma=0$. But $\langle V_{(r,0)} \rangle_\sigma$ may or may not vanish, giving rise to reflecting and sticky boundary conditions respectively. Reflecting boundaries are not special cases of sticky boundaries: as we will deduce from Eq. \eqref{mv}, $\langle V_{(r,0)} \rangle_\sigma\neq 0$ implies constraints that are not obeyed by reflecting boundaries.

Reflecting boundaries are simpler technically, because they are characterised by the 1-point functions $\left<V_P\right>_\sigma$, whereas sticky boundaries would also involve $\langle V_{(r,0)} \rangle_\sigma$. Let us write these 1-point functions, including their dependence on the position of the bulk field on the disc. In fact, we use the conformal equivalence of the disc with the upper half-plane, parametrised by $z\in\mathbb{C}$ with $\Im z>0$. In section \ref{sec:dopf}, we will find
\begin{align}\label{sol}
 \langle V_P(z,\bar z) \rangle_\sigma = \frac{\langle V_P \rangle_\sigma}{|z-\bar z|^{2\Delta(P)}} \qquad \text{with} \qquad  \langle V_P \rangle_\sigma = \sin(4\pi \sigma P) \;.
\end{align}
Now $\sigma\in\mathbb{C}$ is no longer a formal parameter: rather, it parametrises the space of solutions of crossing symmetry equations, in the presence of reflecting boundary conditions. The 1-point structure constant $\langle V_P \rangle_\sigma$ has a particularly simple expression thanks to our choice of normalisation \eqref{eq:diag rescale}.

\subsection{Disc 2-point functions}\label{sec:2pnt function}

The 2-point function $\langle V_{(r_1,s_1)} V_{(r_2,s_2)} \rangle_\sigma$, with $r_1 + r_2 \in \mathbb{N}$, has the bulk channel expansion
\begin{align}
 \left<V_{(r_1,s_1)}V_{(r_2,s_2)}\right>_{\sigma} = \sum_{P\in P_0+\beta^{-1}\mathbb{Z}}   C_{(r_1,s_1)(r_2,s_2)}^P \left<V_P\right>_\sigma
 \begin{tikzpicture}[baseline=(base), scale = .3]
\coordinate (base) at (0, 0);
\draw (-1,2) -- (0,0) -- (4, 0) -- (5,2);
\draw (-1,-2) -- (0,0);
\draw (4,0) -- (5,-2);
\node [above] at (2, 0){$P$};
\node at (5.5, -2.7) {$(r_2,s_2)$};
\node at (5.5, 2.7) {$(r_2,-s_2)$};
\node at (-1.5, -2.7) {$(r_1,s_1)$};
\node at (-1.5, 2.7) {$(r_1, -s_1)$};
 \end{tikzpicture}
 \label{2pt-bulk}
\end{align}
This also applies if one or both fields are diagonal, using the relation \eqref{eq:non-diag to diag}.
We will now describe each part of this expression:
\begin{itemize}
\item\textbf{$r_1 + r_2 \in \mathbb{N}$:} With reflecting boundary conditions, no arcs can end on the boundary of the disc. Since our two fields have $2r_1$ and $2r_2$ legs, we therefore have
\begin{equation}\label{rprn}
  r_1 + r_2 \notin \mathbb{N}\implies  \left<V_{(r_1,s_1)}V_{(r_2,s_2)}\right>_{\sigma} = 0 \;.
\end{equation}

\item\textbf{Diagram \begin{tikzpicture}[baseline=(base), scale = .17]
\coordinate (base) at (0, -0.5);
\draw (-1,2) -- (0,0) -- (4, 0) -- (5,2);
\draw (-1,-2) -- (0,0);
\draw (4,0) -- (5,-2);
 \end{tikzpicture}:} The diagram represents an $s$-channel conformal block. This is a function of the conformal dimensions $(\Delta_{(r_i,s_i)},\Delta_{(r_i,-s_i)})$, the channel field dimension $\Delta(P)$, and the positions of the fields. It can be computed to arbitrary precision using the Zamolodchikov recursion relation \cite{zam84}.

\item\textbf{$C^P_{(r_1,s_1) (r_2,s_2)}$:} The bulk OPE coefficients have been determined previously in \cite{nrj23} and are reviewed in section \ref{sec:scbc}. Explicitly:
\begin{multline}\label{eq:CPr1s1r2s2}
    C^P_{(r_1,s_1) (r_2,s_2)} = \frac{128 P}{\beta^{-1} - \beta} \prod_\pm \frac{\Gamma_\beta(\beta^{\pm1})^2 \Gamma_\beta(\pm2P)}{\Gamma_\beta(2\beta^{\pm1})}\\
    \times \frac{1}{\prod_{\epsilon_1,\epsilon_2,\epsilon_3 = \pm} \Gamma_\beta(\frac{1}{2}(\beta {+} \beta^{-1}) {+} \frac{1}{2}|\epsilon_1 r_1 {+} \epsilon_2 r_2|\beta {+} \frac{1}{2}(\epsilon_1 s_1 {+} \epsilon_2 s_2)\beta^{-1} {+} \epsilon_3 P)} \;.
\end{multline}
If one or both fields are diagonal, the relation \eqref{eq:non-diag to diag} gives
\begin{equation}\label{eq:diag to non-diag OPE}
    C^{P}_{(r_1,s_1)P_2} = \lambda_{P_2} C^{P}_{(r_1,s_1)(0,2P_2\beta)} \;,\quad C^{P}_{P_1 P_2} = \lambda_{P_1} \lambda_{P_2} C^{P}_{(0,2P_1\beta) (0,2P_2\beta)} \;,
\end{equation}
where $\lambda_P$ is defined in \eqref{eq:diag rescale}.

\item\textbf{$\langle V_P \rangle_\sigma$:} The 1-point function defined in \eqref{sol}.

\item\textbf{$P_0$:} This parameter of the bulk CFT determines the weights $w(P_0) = 2\cos(2\pi P_0 \beta)$ of the loops that encircle both fields:
\begin{align}
\begin{tikzpicture}[baseline=(current bounding box.center), scale=2.2]
% smoother, wider-than-tall loop around the point
\draw[thick, red] plot[smooth cycle, tension=0.85] coordinates {
  (0, .45)
  (.42, .34)
  (.62, .08)
  (.50, -.24)
  (.12, -.42)
  (-.32, -.36)
  (-.60, -.08)
  (-.46, .26)
};
\draw (0, .2) ellipse (.8cm and .8cm);
\node at (.3, .55) {$w(P_0)$};
\node at (0.7, 0.8) {$\sigma$};
% central point and label
\draw (0., 0.2) node[fill, circle, minimum size = 1.2mm, inner sep = 0]{};
\node[left] at (.12, -.11) {$V_{(r_1,s_1)}$};
\draw (-0.5, -0.1) node[fill, circle, minimum size = 1.2mm, inner sep = 0]{};
\node[left] at (.63, 0.18) {$V_{(r_2,s_2)}$};
\end{tikzpicture}
\end{align}
The parameter $P_0$ determines the bulk channel spectrum $\{V_P | P \in P_0 + \beta^{-1}\mathbb{Z}, P_0 \in \mathbb{C}\}$, which is one interchiral family of diagonal fields that share the same weight $w(P_0)$ \cite{rib24}.
\end{itemize}

\subsection{Boundary spectrum}\label{sec:bndry fields}

We further subdivide reflecting boundary conditions into two cases: continuous and discrete reflecting boundary conditions, depending on the continuous or discrete nature of the boundary spectrum:
\begin{itemize}
    \item \textbf{Continuous:} $\sigma \in \mathbb{C}$ such that $\sigma \neq P_{(0,S)}$ for $S\in\mathbb{Z}_{\neq0}$.

    \item\textbf{Discrete:} $\sigma = P_{(0,S)}$ with $S\in\mathbb{N}^*$.
\end{itemize}
Correlation functions have a definite parity in $\sigma$, so we can omit the values $\sigma = -P_{(0,S)}$ with $S\in\mathbb{N}^*$.

A boundary field is characterised by a conformal dimension $\Delta$, and the boundary conditions on each side. In general, these two boundary conditions can differ. In this article, we do not consider boundary-condition-changing fields, and we do not include boundary parameters as labels of boundary fields. However, in the following table, we specify on which type of boundary condition the fields can live. There are two types of boundary fields:
\begin{align}\label{tab:bndry fields}
\renewcommand{\arraystretch}{1.5}
 \begin{array}{|r|l|l|l|l|l|}
  \hline
  \text{Name} & \text{Notation} & \text{Conditions} & P & \text{Boundary condition}
  \\
  \hline \hline
  \text{Degenerate} &  \psi^d_{\langle r,s\rangle} &  r\in\mathbb{N}^* \;,\; s\in2\mathbb{N}^*{-}1 & P_{(r,s)} & \sigma = P_{(0,S)} \;,\; S \geq \frac{s+1}{2}
  \\
  \hline
  \text{Generic} & \psi_P & P\in\mathbb{C} & P & \sigma \in \mathbb{C}\setminus\{\pm P_{(0,S)} | S\in \mathbb{N}^*\}
  \\
  \hline
 \end{array}
\end{align}
Degenerate boundary fields have a geometric interpretation in loop models \cite{car05}. A degenerate boundary field $\psi^d_{\langle r,s\rangle}$ emits $r-1$ arcs from a point on the boundary:
\begin{align}
\begin{tikzpicture}[baseline=(current  bounding  box.center), scale = 1.4]
 \node[below] at (0, -.1) {$\psi^d_{\langle r,s\rangle}$};
 \node[below] at (1, 0) {$\sigma$};
 \node[below] at (-1, 0) {$\sigma$};
 \begin{scope}[rotate = 90]
 \draw[thick, red] plot [smooth] coordinates {(0, 0)(.2, .7)(.4, 1.2)};
\node at (.5, 1.3) {$\scriptstyle{1}$};
\draw[thick, red] plot [smooth] coordinates {(0, 0)(.3, .6)(.6, 1)};
\node at (.7, 1.1) {$\scriptstyle{2}$};
\draw[thick, red] plot [smooth] coordinates {(0, 0)(.4, .5)(.8, .8)};
\node at (.9, .9) {$\scriptstyle{3}$};
\draw[thick, red] plot [smooth] coordinates {(0, 0)(.5, -.3)(1.1, -.5)};
\node at (1.3, -.5) {$\scriptstyle{r-1}$};
\draw[dashed] plot [smooth] coordinates {(.6, .45)(.75, .1)(.7, -.25)};
\end{scope}
 \draw (-1.6, 0) -- (1.6, 0);
 \draw (0, 0) node[fill, circle, minimum size = 1.6mm, inner sep = 0]{};
\end{tikzpicture}
\end{align}
If the boundary is discrete, the boundary channel expansion of a 2-point function is a discrete sum:
\begin{align}
 \left<V_{(r_1,s_1)}V_{(r_2,s_2)}\right>_{P_{(0,S)}} = \sum_{\psi^d_{\langle r,s\rangle}\in \mathcal{S}_S^{r_1,r_2}} D_{\langle r,s\rangle}
 \begin{tikzpicture}[baseline=(base), scale = .35]
\coordinate (base) at (0, 2);
\draw (2, -1) -- (0,0) -- (0, 4) -- (2,5);
\draw (-2, -1) -- (0,0);
\draw (0,4) -- (-2,5);
\node [left] at (0,2){$\langle r,s\rangle$};
\node at (2.5, -1.7) {$(r_2,s_2)$};
\node at (2.5, 5.7) {$(r_2,-s_2)$};
\node at (-2.5, -1.7) {$(r_1,s_1)$};
\node at (-2.5, 5.7) {$(r_1, -s_1)$};
 \end{tikzpicture}
 \label{2pt-bdy}
\end{align}
Let us describe each piece of this expansion:
\begin{itemize}
\item \textbf{Diagrams  \begin{tikzpicture}[baseline=(base), scale = .17]
\coordinate (base) at (0, 1.4);
\draw (2, -1) -- (0,0) -- (0, 4) -- (2,5);
\draw (-2, -1) -- (0,0);
\draw (0,4) -- (-2,5);
 \end{tikzpicture}:} The diagrams represent $t$-channel conformal blocks. Again these can be evaluated using the Zamolodchikov recursion relation \cite{zam84}.

\item \textbf{$\mathcal{S}_S^{r_1,r_2}$:} The boundary spectrum, derived in section \ref{sec:bou}, is the following infinite set of degenerate boundary fields
\begin{align}
 \mathcal{S}_S^{r_1,r_2} = \left\{\psi^d_{\langle r,s\rangle}\right\}_{\substack{r\in 2\max(r_1,r_2)+1+2\mathbb{N} \\ s = 1, 3, \dots, 2S-1}} \;.
 \label{s12}
\end{align}
\item\textbf{$D_{\langle r,s\rangle}$:} The structure constants $D_{\langle r,s\rangle}$ can be computed numerically. They depend on the Kac indices $(r_1,s_1),(r_2,s_2)$ of bulk fields, on the indices $\langle r,s\rangle$ of channel fields, and on $P_0$ from the bulk channel decomposition \eqref{2pt-bulk}. We leave the analytic determination of $D_{\langle r,s\rangle}$ to future work.
\end{itemize}
However, in the case of a 2-point function of diagonal fields $\left<V_{P_1}V_{P_2}\right>_{P_{(0,S)}}$, we have an analytic expression for $D_{\langle1,1\rangle}$
\begin{align}
 D_{\langle 1,1\rangle} =  \frac{\sin(2\pi\beta^{-1}SP_1)\sin(2\pi\beta^{-1}SP_2)}{(-1)^{S}\sin(\pi\beta^{-2}S)}\ .
 \label{d11}
\end{align}
In section \ref{sec:num checks}, we will derive this expression and test it numerically. We will also check that the bulk- \eqref{2pt-bulk} and boundary-channel \eqref{2pt-bdy} expansions are compatible, in other words, that our disc 2-point function obeys crossing symmetry. However, we are technically unable to perform this check in the case of continuous boundary conditions, because the sum in \eqref{2pt-bdy} would be an integral.

\section{Analytic bootstrap}\label{sec:analytic bootstrap}

In this section we will derive the results that we announced in section \ref{sec:main results}.

We will begin by reviewing structure constants of the bulk CFT, and showing that the normalisation \eqref{eq:diag rescale} makes some of them simpler.

Then we will analyse analytic bootstrap constraints for disc 1-point functions. This will lead us to the reflecting and sticky boundary conditions, and allow us to determine the disc 1-point functions. We will also distinguish continuous and discrete reflecting boundary conditions based on the behaviour of the disc 1-point functions as $P\rightarrow P_{(r,s)}$ for $r,s\in\mathbb{N}^*$.

Using the bulk 2-point functions and the disc 1-point functions, we will compute annulus partition functions in the bulk channel. A modular transformation will then give us access to the boundary channel, and therefore to boundary spectra. We will thus re-derive which boundary conditions are discrete and which are continuous.

Finally in section \ref{sec:bou} we will show that the boundary channel expansion of a given disc 2-point function involves a subset of the relevant boundary spectrum.

\subsection{Structure constants of the bulk CFT}\label{sec:scbc}

Before looking at disc correlators we first make a brief detour via the bulk CFT. We will also introduce shift equations, which are discussed in more detail in \cite{rib24}.

The existence of bulk degenerate fields allows us to use the analytic bootstrap method. This method is based on solving crossing symmetry equations for correlation functions that involve the simplest non-trivial degenerate field $V^d_{\langle 1,2\rangle}$, whose OPEs with non-degenerate fields are
\begin{align}\label{eq:aux OPE}
 V^d_{\langle 1,2\rangle}V_P \sim \sum_\pm C^\pm_P V_{P\pm \frac{1}{2\beta}} \quad , \quad V^d_{\langle 1,2\rangle} V_{(r,s)} \sim \sum_\pm C^\pm_{(r,s)} V_{(r,s\pm 1)}\ .
\end{align}
In these OPEs we have been schematic and written only primary fields and omitted descendants. The $C^\pm_P$ and $C^\pm_{(r,s)}$ are OPE coefficients. The resulting analytic bootstrap constraints determine how correlation functions behave under $P\to P+\beta^{-1}$ or $s\to s+2$.

The normalisation \eqref{eq:diag rescale} simplifies the shift equation for the sphere 2-point structure constant $B_P\propto \left< V_P V_P \right>^\text{sphere}$ which appears in the annulus partition function in section \ref{sec:modular}. Consider the $s$- and $t$-channel decompositions of the following sphere four-point function:
\begin{align}
 \left\langle V_{\langle1,2\rangle} V_{\langle1,2\rangle} V_P  V_P \right\rangle^\text{sphere}
 &= \frac{B_{\langle 1,2\rangle}B_P}{B_{\langle 1,1\rangle}} \left|
 \begin{tikzpicture}[baseline=(base), scale = .3]
\coordinate (base) at (0, 0);
\draw (-1,2) -- (0,0) -- (4, 0) -- (5,2);
\draw (-1,-2) -- (0,0);
\draw (4,0) -- (5,-2);
\node [above] at (2, 0){$\scriptstyle{\langle1,1\rangle}$};
\node at (5.5, -2.7) {$\scriptstyle{P}$};
\node at (5.5, 2.7) {$\scriptstyle{P}$};
\node at (-1.5, -2.7) {$\scriptstyle{\langle1,2\rangle}$};
\node at (-1.5, 2.7) {$\scriptstyle{\langle1,2\rangle}$};
 \end{tikzpicture}
\right|^2 + \tilde{D}_{\langle 1,3\rangle} \left|
 \begin{tikzpicture}[baseline=(base), scale = .3]
\coordinate (base) at (0, 0);
\draw (-1,2) -- (0,0) -- (4, 0) -- (5,2);
\draw (-1,-2) -- (0,0);
\draw (4,0) -- (5,-2);
\node [above] at (2, 0){$\scriptstyle{\langle1,3\rangle}$};
\node at (5.5, -2.7) {$\scriptstyle{P}$};
\node at (5.5, 2.7) {$\scriptstyle{P}$};
\node at (-1.5, -2.7) {$\scriptstyle{\langle1,2\rangle}$};
\node at (-1.5, 2.7) {$\scriptstyle{\langle1,2\rangle}$};
 \end{tikzpicture}
\right|^2\ ,
\nonumber
\\
&= \sum_\pm \left(C_P^\pm\right)^2 B_{P\pm \frac{1}{2\beta}}
\left|\begin{tikzpicture}[baseline=(base), scale = .3]
\coordinate (base) at (0, 2);
\draw (2, -1) -- (0,0) -- (0, 4) -- (2,5);
\draw (-2, -1) -- (0,0);
\draw (0,4) -- (-2,5);
\node [right] at (0,2){$\scriptstyle{P\pm\frac{1}{2\beta}}$};
\node at (2.5, -1.7) {$\scriptstyle{P}$};
\node at (2.5, 5.7) {$\scriptstyle{P}$};
\node at (-2.5, -1.7) {$\scriptstyle{\langle1,2\rangle}$};
\node at (-2.5, 5.7) {$\scriptstyle{\langle1,2\rangle}$};
 \end{tikzpicture}\right|^2\  .
 \label{4sphere}
\end{align}
In addition to $B_P$, these decompositions involve the degenerate OPE coefficients $C_P^\pm$ \eqref{eq:aux OPE} and a structure constant $\tilde{D}_{\langle 1,3\rangle}$. The diagrams denote $s$- and $t$-channel Virasoro conformal blocks, and the modulus squared means that we take products of left-moving blocks (functions of the cross-ratio $z$ of the 4 fields' positions) with right-moving blocks (functions of $\bar z$).
The $s$- and $t$-channel blocks are linearly related,
\begin{equation}\label{eq:diagF}
\begin{tikzpicture}[baseline=(base), scale = .3]
\coordinate (base) at (0, 2);
\draw (2, -1) -- (0,0) -- (0, 4) -- (2,5);
\draw (-2, -1) -- (0,0);
\draw (0,4) -- (-2,5);
\node [right] at (0,2){$\scriptstyle{P\pm\frac{1}{2\beta}}$};
\node at (2.5, -1.7) {$\scriptstyle{P}$};
\node at (2.5, 5.7) {$\scriptstyle{P}$};
\node at (-2.5, -1.7) {$\scriptstyle{\langle1,2\rangle}$};
\node at (-2.5, 5.7) {$\scriptstyle{\langle1,2\rangle}$};
 \end{tikzpicture}
 =
 f^\pm_P
\begin{tikzpicture}[baseline=(base), scale = .3]
\coordinate (base) at (0, 0);
\draw (-1,2) -- (0,0) -- (4, 0) -- (5,2);
\draw (-1,-2) -- (0,0);
\draw (4,0) -- (5,-2);
\node [above] at (2, 0){$\scriptstyle{\langle1,1\rangle}$};
\node at (5.5, -2.7) {$\scriptstyle{P}$};
\node at (5.5, 2.7) {$\scriptstyle{P}$};
\node at (-1.5, -2.7) {$\scriptstyle{\langle1,2\rangle}$};
\node at (-1.5, 2.7) {$\scriptstyle{\langle1,2\rangle}$};
 \end{tikzpicture}
 + \tilde{f}^\pm_P
 \begin{tikzpicture}[baseline=(base), scale = .3]
\coordinate (base) at (0, 0);
\draw (-1,2) -- (0,0) -- (4, 0) -- (5,2);
\draw (-1,-2) -- (0,0);
\draw (4,0) -- (5,-2);
\node [above] at (2, 0){$\scriptstyle{\langle1,3\rangle}$};
\node at (5.5, -2.7) {$\scriptstyle{P}$};
\node at (5.5, 2.7) {$\scriptstyle{P}$};
\node at (-1.5, -2.7) {$\scriptstyle{\langle1,2\rangle}$};
\node at (-1.5, 2.7) {$\scriptstyle{\langle1,2\rangle}$};
 \end{tikzpicture}
\end{equation}
where $f^\pm_P,\tilde{f}^\pm_P$ are fusion matrix elements which can be found in \cite{rib24}. We will only need the explicit form of
\begin{equation}
    f_P^\pm = \frac{\Gamma(2\beta^{-2} - 1) \Gamma(1 \pm 2P \beta^{-1})}{\Gamma(\beta^{-2}) \Gamma(\beta^{-2} \pm 2P \beta^{-1})} \;.
\end{equation}
Inserting this relation into \eqref{4sphere}, and considering the coefficient of $\Bigg|
 \begin{tikzpicture}[baseline=(base), scale = .2]
\coordinate (base) at (0, -.5);
\draw (-1,2) -- (0,0) -- (4, 0) -- (5,2);
\draw (-1,-2) -- (0,0);
\draw (4,0) -- (5,-2);
\node [above] at (2, 0){$\scriptstyle{\langle1,1\rangle}$};
\node at (5.5, -2.7) {$\scriptstyle{P}$};
\node at (5.5, 2.7) {$\scriptstyle{P}$};
\node at (-1.5, -2.7) {$\scriptstyle{\langle1,2\rangle}$};
\node at (-1.5, 2.7) {$\scriptstyle{\langle1,2\rangle}$};
 \end{tikzpicture}
\Bigg|^2$, we obtain a shift equation for $B_P$,
\begin{align}
 \frac{B_{\langle 1,2\rangle}B_P}{B_{\langle 1,1\rangle}} = \sum_\pm \left(C_P^\pm f_P^\pm\right)^2 B_{P\pm \frac{1}{2\beta}}\ .
\end{align}
We slightly simplify this equation by normalising the fields such that $B_{\langle 1,1\rangle}=1$. This is a natural choice of normalisation since $V^d_{\langle 1,1\rangle}$ is the bulk identity field. A further simplification occurs if we choose field normalisations such that
\begin{equation}\label{eq:diag aux}
    C^\pm_P = (f^\pm_P)^{-1} \;.
\end{equation}
And indeed, this identity is obeyed if we start from the normalisations of \cite{rib24}, and perform the field renormalisation $V_k = \lambda_k \tilde V_k$ so that $C^\pm_P = \frac{\lambda_{\langle 1,2\rangle}\lambda_P}{\lambda_{P\pm\frac{1}{2\beta}}}\tilde C_P^\pm$, with
\begin{equation}\label{eq:scaling}
    \lambda_{\langle1,2\rangle} = -\frac{4\sqrt{\pi} (4\beta)^{-4\beta^{-2}}}{\Gamma(\beta^{-2} - \frac{1}{2})} \;,
\end{equation}
and $\lambda_P$ given in \eqref{eq:diag rescale}. Here the tildes denote fields and OPE coefficients in the normalisation of \cite{rib24}.

Since $\lambda_P$ is invariant under $\beta\to \beta^{-1}$, the shift equation from the four-point function $\left\langle V^d_{\langle 2,1\rangle} V^d_{\langle 2,1\rangle} V_P  V_P \right\rangle^\text{sphere}$ also simplifies (assuming $V^d_{\langle 2,1\rangle}$ exists), and the two shift equations  read
\begin{align}
B_{\langle 2,1\rangle}B_P = \sum_\pm  B_{P\pm \frac{\beta}{2}}
\quad , \quad
 B_{\langle 1,2\rangle}B_P = \sum_\pm  B_{P\pm \frac{1}{2\beta}} \ .
\end{align}
The solution of these shift equations is
\begin{align}
 B_P = -\prod_\pm \frac{\sin\left(2\pi\beta^{\pm 1}P\right)}{\sin\left(\pi\beta^{\pm 2}\right)}\quad \text{with} \quad \left\{
 \renewcommand{\arraystretch}{1.4}
 \begin{array}{l} B_{\langle 2,1\rangle}= -2\cos(\pi\beta^2)\ , \\ B_{\langle 1,2\rangle}=-2\cos(\pi\beta^{-2}) \ . \end{array}\right.
 \label{bp}
\end{align}
This solution is unique only up to a $P$-independent prefactor, which we have chosen such that $B_{P_{( 1,1)}}=1$, consistently with the identification $B_{\langle 1,1\rangle}=B_{P_{(1,1)}}$. Then we also have $B_{\langle 2,1\rangle}=B_{P_{(2,1)}}$ and $B_{\langle 1,2\rangle}=B_{P_{(1,2)}}$.

Our expression for the bulk 2-point structure constant of diagonal fields $B_P$ is valid in critical loop models, and also in Liouville theory with $c\leq 1$. And indeed it can alternatively be derived from the known structure constants of these theories, by applying a field renormalisation such that our normalisation condition \eqref{eq:diag aux} is obeyed.

We also want the bulk OPE coefficient $C^P_{(r_1,s_1)(r_2,s_2)}$ that appears in \eqref{2pt-bulk}. While bulk OPEs in critical loop models are not known in general, let us argue that enough is known for computing our disc 2-point function. First, it is known that $r$ is conserved modulo integers \cite{rib24}, therefore,
\begin{align}
 \left<V_{(r_1,s_1)}V_{(r_2,s_2)}\right>_\sigma \neq 0 \implies r_1+r_2\in \mathbb{N}\ .
\end{align}
In CFT, we generally expect that OPE coefficients are related to 3-point and 2-point structure constants by the simple relation $C_{ij}^k = \frac{C_{ijk}}{B_k}$. However, in critical loop models, this relation does not hold in general. We know the 2-point and 3-point structure constants $B_k,C_{ijk}$ \cite{jnrr25}, but they are not enough for computing sphere four-point functions: extra factors are needed, which are polynomial in loop weights \cite{nrj23}. Nevertheless, these extra factors were found to be trivial whenever the channel field is diagonal.
In the following OPEs we assume $r_1+r_2\in \mathbb{N}$. Under this condition, diagonal fields can appear in the OPE $V_{(r_1,s_1)}V_{(r_2,s_2)}$. Such fields form a discrete infinite family $\{V_P\}_{P\in P_0+\beta^{-1}\mathbb{Z}}$, characterised by the momentum $P_0$. We thus expect an OPE of the type
\begin{align}
 V_{(r_1,s_1)}V_{(r_2,s_2)} \sim \sum_{P\in P_0+\beta^{-1}\mathbb{Z}} C^P_{(r_1,s_1)(r_2,s_2)} V_P + \sum_{r\in\mathbb{N}^*}\sum_{s\in\frac{1}{r}\mathbb{Z}} C^{(r,s)}_{(r_1,s_1)(r_2,s_2)} V_{(r,s)}\ ,
 \label{bulkOPE}
\end{align}
where we write primary fields while omitting descendant fields. 
Since we are considering reflecting boundary conditions, the non-diagonal OPE coefficients $C^{(r,s)}_{(r_1,s_1)(r_2,s_2)}$ do not contribute to the disc 2-point function. We therefore only want
\begin{align}
 C^P_{(r_1,s_1)(r_2,s_2)}  = \frac{C_{(r_1,s_1)(r_2,s_2)P}}{B_P}= \lambda_P\frac{\tilde{C}_{(r_1,s_1)(r_2,s_2)P}}{B_P} \ ,
\end{align}
where $\lambda_P$ is the renormalisation factor \eqref{eq:diag rescale}, $B_P$ is the 2-point structure constant in our normalisations \eqref{bp}, and $\tilde{C}_{(r_1,s_1)(r_2,s_2)P}$ is a 3-point structure constant in the normalisation of \cite{jnrr25}, namely
\begin{align}
  \tilde{C}_{(r_1,s_1)(r_2,s_2)(r_3,s_3)} =\prod_{\epsilon_1,\epsilon_2,\epsilon_3=\pm} \Gamma_\beta^{-1} \left(\tfrac{\beta+\beta^{-1}}{2} + \tfrac{\beta}{2}\left|\textstyle{\sum_i} \epsilon_ir_i\right| + \tfrac{\beta^{-1}}{2}\textstyle{\sum_i} \epsilon_is_i\right)\ .
  \label{cref}
\end{align}
and then
\begin{equation}
    \tilde{C}_{(r_1,s_1)(r_2,s_2)P} = \tilde{C}_{(r_1,s_1)(r_2,s_2)(0,2P\beta)} \;.
\end{equation}
The explicit evaluation of $C^P_{(r_1,s_1)(r_2,s_2)}$ leads to the formula \eqref{eq:CPr1s1r2s2}.

\subsection{Reflecting and sticky boundary conditions}\label{sec:dand}

Consider the bulk-to-boundary OPE of the degenerate field $V^d_{\langle 1,2\rangle}$ as it approaches the boundary with boundary condition $\sigma$
\begin{equation}\label{eq:V12 to bndry}
    V^d_{\langle 1,2\rangle} \to \mu_\sigma \psi^d_{\langle1,1\rangle} + \tilde{\mu}_\sigma \psi^d_{\langle1,3\rangle} \;.
\end{equation}
The degenerate boundary fields in the OPE were introduced in table \eqref{tab:bndry fields} and $\mu_\sigma$ and $\tilde \mu_\sigma$ are the OPE coefficients.

The vanishing of the singular vector of $V^d_{\langle 1,2\rangle}$ implies that its bulk-to-boundary OPE produces boundary fields with momenta $P_{(1,1)}$ and $P_{(1,3)}$: this can be seen using Ward identities or equivalently BPZ equations. But does this OPE produce the degenerate fields $\psi^d_{\langle 1,1\rangle}$ and $\psi^d_{\langle 1,3\rangle}$, or the generic fields $\psi_{P_{(1,1)}}$, $\psi_{P_{(1,3)}}$ from table \eqref{tab:bndry fields}? In the bulk CFT, the analogous question
about the OPE $V^d_{\langle 1,2\rangle}V^d_{\langle 1,2\rangle}$ is answered using OPE associativity, and the fact that a field is degenerate if and only if its OPE with any other primary field produces finitely many primary fields \cite{rib24}. The same argument would show that the set of boundary degenerate fields is closed under OPEs, but it does not work for bulk-to-boundary OPEs. In that case, we can resort to the pedestrian argument of solving the BPZ equation for the disc correlation function $\left<V^d_{\langle 1,2\rangle}\psi_{p_1}\psi_{p_2}\right>_\sigma$, where $\psi_{p}$ is a primary boundary field of momentum $p\in\mathbb{C}$. This correlator on the disc satisfies the same Ward identities as a sphere four-point function of four holomorphic fields. Two of the fields in the sphere four-point function are degenerate; hence, we have two second-order BPZ equations. We find that the compatibility of these two BPZ equations implies
\begin{align}
 \prod_\pm (p_1\pm p_2)\prod_{\pm,\pm}(p_1\pm p_2\pm \beta^{-1})=0\ .
\end{align}
This is equivalent to $\left<\psi^d_{\langle 1,1\rangle}\psi_{p_1}\psi_{p_2}\right>_\sigma\neq 0$ or $\left<\psi^d_{\langle 1,3\rangle}\psi_{p_1}\psi_{p_2}\right>_\sigma\neq 0$, which shows that the bulk-to-boundary OPE of $V^d_{\langle 1,2\rangle}$ produces degenerate fields.

We now consider the disc 2-point function $\langle V_{\langle1,2\rangle}^d V_{(r,s)} \rangle$, and decompose it in both the bulk and boundary channels, according to the bulk and bulk-to-boundary OPEs \eqref{eq:aux OPE} and \eqref{eq:V12 to bndry}. Schematically,
\begin{align}
\begin{tikzpicture}[baseline=(current  bounding  box.center), scale=1.2, every node/.style={scale=1.1}]
        \draw (1.5,0) arc [start angle=0,end angle=3600,x radius=1.5,y radius=1.5];
        \node at (-0.6,0.4) {$V_{\langle1,2\rangle}^d$};
        \node at (0.6,0.4) {$V_{(r,s)}$};
        \draw[ultra thick, blue, -latex] (-0.5,0.11) -- (-0.03,-0.3);
        \draw[ultra thick, blue, -latex] (0.5,0.11) -- (0.03,-0.3);
        \node[blue] at (0,-0.5) {$V_{(r,s\pm1)}$};
        \node at (0, -2){bulk channel};
\end{tikzpicture}
\hspace{2cm}
\begin{tikzpicture}[baseline=(current  bounding  box.center), scale=1.2, every node/.style={scale=1.1}]
        \draw (1.5,0) arc [start angle=0,end angle=3600,x radius=1.5,y radius=1.5];
        \node at (-0.55,0) {$V_{\langle1,2\rangle}^d$};
        \node at (0.55,0) {$V_{(r,s)}$};
        \draw[ultra thick, blue, -latex] (-.95,0) -- (-1.45,0);
        \node[left, blue] at (-1.5,0) {$\psi_{\langle1,1\rangle}^d,\psi_{\langle1,3\rangle}^d$};
        \node at (0, -2){boundary channel};
\end{tikzpicture}
\end{align}
This leads to two decompositions of $\langle V_{\langle1,2\rangle}^d V_{(r,s)} \rangle$ into conformal blocks,
\begin{align}\label{eq:nondiag blocks}
    \left\langle V_{\langle1,2\rangle}^d V_{(r,s)} \right\rangle_\sigma
  &=
 \mu_\sigma \langle V_{(r,s)}\rangle_\sigma
 \begin{tikzpicture}[baseline=(base), scale = .3]
\coordinate (base) at (0, 0);
\draw (-1,2) -- (0,0) -- (4, 0) -- (5,2);
\draw (-1,-2) -- (0,0);
\draw (4,0) -- (5,-2);
\node [above] at (2, 0){$\scriptstyle{\langle1,1\rangle}$};
\node at (5.5, -2.7) {$\scriptstyle{(r,-s)}$};
\node at (5.5, 2.7) {$\scriptstyle{(r,s)}$};
\node at (-1.5, -2.7) {$\scriptstyle{\langle1,2\rangle}$};
\node at (-1.5, 2.7) {$\scriptstyle{\langle1,2\rangle}$};
 \end{tikzpicture} +\tilde{d}_{\langle 1,3\rangle}
 \begin{tikzpicture}[baseline=(base), scale = .3]
\coordinate (base) at (0, 0);
\draw (-1,2) -- (0,0) -- (4, 0) -- (5,2);
\draw (-1,-2) -- (0,0);
\draw (4,0) -- (5,-2);
\node [above] at (2, 0){$\scriptstyle{\langle1,3\rangle}$};
\node at (5.5, -2.7) {$\scriptstyle{(r,-s)}$};
\node at (5.5, 2.7) {$\scriptstyle{(r,s)}$};
\node at (-1.5, -2.7) {$\scriptstyle{\langle1,2\rangle}$};
\node at (-1.5, 2.7) {$\scriptstyle{\langle1,2\rangle}$};
 \end{tikzpicture}
 \ ,
 \nonumber
\\
 &=
    \sum_\pm C^\pm_{(r,s)} \langle V_{(r,s\pm1)} \rangle_\sigma
    \begin{tikzpicture}[baseline=(base), scale = .3]
\coordinate (base) at (0, 2);
\draw (2, -1) -- (0,0) -- (0, 4) -- (2,5);
\draw (-2, -1) -- (0,0);
\draw (0,4) -- (-2,5);
\node [right] at (0,2){$\scriptstyle{(r,s\pm1)}$};
\node at (2.5, -1.7) {$\scriptstyle{(r,-s)}$};
\node at (2.5, 5.7) {$\scriptstyle{(r,s)}$};
\node at (-2.5, -1.7) {$\scriptstyle{\langle1,2\rangle}$};
\node at (-2.5, 5.7) {$\scriptstyle{\langle1,2\rangle}$};
 \end{tikzpicture}
 \ ,
\end{align}
where $\mu_\sigma$ is defined in \eqref{eq:V12 to bndry}, and $\tilde{d}_{\langle 1,3\rangle}$ is a boundary channel 2-point structure constant. The conformal blocks in the two channels are related by a linear transformation of the type
\begin{equation}
\begin{tikzpicture}[baseline=(base), scale = .3]
\coordinate (base) at (0, 2);
\draw (2, -1) -- (0,0) -- (0, 4) -- (2,5);
\draw (-2, -1) -- (0,0);
\draw (0,4) -- (-2,5);
\node [right] at (0,2){$\scriptstyle{(r,s\pm1)}$};
\node at (2.5, -1.7) {$\scriptstyle{(r,-s)}$};
\node at (2.5, 5.7) {$\scriptstyle{(r,s)}$};
\node at (-2.5, -1.7) {$\scriptstyle{\langle1,2\rangle}$};
\node at (-2.5, 5.7) {$\scriptstyle{\langle1,2\rangle}$};
 \end{tikzpicture}
 =
 f^\pm_{(r,s)}
\begin{tikzpicture}[baseline=(base), scale = .3]
\coordinate (base) at (0, 0);
\draw (-1,2) -- (0,0) -- (4, 0) -- (5,2);
\draw (-1,-2) -- (0,0);
\draw (4,0) -- (5,-2);
\node [above] at (2, 0){$\scriptstyle{\langle1,1\rangle}$};
\node at (5.5, -2.7) {$\scriptstyle{(r,-s)}$};
\node at (5.5, 2.7) {$\scriptstyle{(r,s)}$};
\node at (-1.5, -2.7) {$\scriptstyle{\langle1,2\rangle}$};
\node at (-1.5, 2.7) {$\scriptstyle{\langle1,2\rangle}$};
 \end{tikzpicture}
 +
 \tilde{f}^\pm_{(r,s)}
\begin{tikzpicture}[baseline=(base), scale = .3]
\coordinate (base) at (0, 0);
\draw (-1,2) -- (0,0) -- (4, 0) -- (5,2);
\draw (-1,-2) -- (0,0);
\draw (4,0) -- (5,-2);
\node [above] at (2, 0){$\scriptstyle{\langle1,3\rangle}$};
\node at (5.5, -2.7) {$\scriptstyle{(r,-s)}$};
\node at (5.5, 2.7) {$\scriptstyle{(r,s)}$};
\node at (-1.5, -2.7) {$\scriptstyle{\langle1,2\rangle}$};
\node at (-1.5, 2.7) {$\scriptstyle{\langle1,2\rangle}$};
 \end{tikzpicture}
 \ ,
\end{equation}
where the fusion matrix elements $f_{(r,s)}^\pm$ are \cite{rib24}
\begin{equation}\label{eq:nondiagF}
    f_{(r,s)}^\pm =  \frac{\Gamma(2\beta^{-2} - 1) \Gamma(1 \mp (r-s\beta^{-2}))}{\Gamma(\beta^{-2} \mp r) \Gamma((1\pm s)\beta^{-2})}\;.
\end{equation}
Projecting the crossing symmetry equation \eqref{eq:nondiag blocks} on the boundary identity channel, we obtain the shift equation
\begin{align}
 \mu_\sigma\left<V_{(r,s)}\right>_\sigma =\sum_\pm C^\pm_{(r,s)}f^\pm_{(r,s)} \left<V_{(r,s\pm 1)}\right>_\sigma \ ,
 \label{mvrs}
\end{align}
for the disc 1-point function. By conformal invariance, $\left<V_{(r,s)}\right>_\sigma$ must vanish unless the conformal spin $rs$ vanishes, so that
\begin{align}
 \left<V_{(r,s\neq 0)}\right>_\sigma = 0 \ .
 \label{conf}
\end{align}
Combining this constraint with the case $s=0$ of Eq. \eqref{mvrs}, we obtain
\begin{align}
 \mu_\sigma \left<V_{(r,0)}\right>_\sigma = 0 \ .
 \label{mv}
\end{align}
This leads us to distinguish 2 types of boundary conditions:
\begin{itemize}
 \item \textbf{Reflecting boundary conditions:} $\left<V_{(r,0)}\right>_\sigma=0$: only diagonal or degenerate fields ($V_P$ or $V^d_{\langle 1,s\rangle}$) have non-zero disc 1-point functions.
 \item \textbf{Sticky boundary conditions:} $\left<V_{(r,0)}\right>_\sigma \neq 0$, therefore, we must have $\mu_\sigma=0$. In this case spinless non-diagonal fields $V_{(r,0)}$ can also have non-zero disc 1-point functions.
\end{itemize}
We have seen how the special case $s=0$ of the shift equation leads to the distinction between reflecting and sticky boundary conditions. Sticky boundary conditions obey $\left<V_{(r,0)}\right>\neq 0$ and $\left<V_{(r,s\neq 0)}\right>=0$; let us check that this is compatible with the general case of the shift equation. For $s\notin\{-1,0,1\}$, all three 1-point functions in \eqref{mvrs} vanish, so \eqref{mvrs} contains no information in these cases. The case $s=0$ is \eqref{mv}. Finally, for $s=\mp 1$ we obtain the non-trivial relation
\begin{equation}\label{eq:nondiag check}
    0 = C^\pm_{(r,\mp1)} f^\pm_{(r,\mp1)} \langle V_{(r,0)} \rangle_\sigma \;.
\end{equation}
From Eq. \eqref{eq:nondiagF} we see that $f^\pm_{(r,\mp1)} = 0$, so this equation is obeyed, and does not constrain $\left<V_{(r,0)}\right>_\sigma$.

\subsection{Disc 1-point functions}\label{sec:dopf}

Let us derive shift equations for the disc 1-point functions of diagonal fields $\left<V_P\right>_\sigma$. We decompose the disc 2-point function $\langle V_{\langle1,2\rangle}^d V_P \rangle_\sigma$ in both the bulk and boundary channels, just like we did for $\langle V_{\langle1,2\rangle}^d V_{(r,s)} \rangle_\sigma$ in \eqref{eq:nondiag blocks}. By the folding trick \cite{Northe:2024tnm}, the conformal blocks that appear are the same as for the four-point function $\left\langle V^d_{\langle1,2\rangle} V^d_{\langle1,2\rangle} V_P  V_P \right\rangle^\text{sphere}$ in \eqref{4sphere}.

Crossing symmetry of the disc 2-point function $\left<V^d_{\langle 1,2\rangle}V_P\right>_\sigma$ implies the following shift equation
\begin{align}
 \mu_\sigma \left<V_P\right>_\sigma = \sum_\pm \left<V_{P\pm \frac{1}{2\beta}}\right>_\sigma\ .
 \label{mvsv}
\end{align}
This is similar to the shift equation \eqref{mvrs} for $\left<V_{(r,s)}\right>_\sigma$, but we now have trivial coefficients on the right-hand side, due to our choice of a field normalisation that obeys Eq. \eqref{eq:diag aux}.

Let us now solve the shift equation \eqref{mvsv}, together with the parity constraint $\left<V_P\right>_\sigma = -\left<V_{-P}\right>_\sigma$ which follows from \eqref{odd}. We choose to parametrise $\mu_\sigma$ as
\begin{equation}
    \mu_\sigma = 2\cos(2\pi \sigma \beta^{-1}) \;.
\end{equation}
The general solution is
\begin{align}
 \left<V_P\right>_\sigma = \sum_{\sigma'\in\sigma+\beta\mathbb{Z}} c_{\sigma'} \sin(4\pi \sigma' P) \quad \text{with} \quad c_{\sigma'}\in\mathbb{C}\ .
 \label{sisz}
\end{align}
The existence of such a large space of solutions is due to the absence of a second degenerate field $V^d_{\langle 2,1\rangle}$, which would give rise to a second shift equation:
\begin{align}
 \bar{\mu}_\sigma\left<V_P\right>_\sigma = \sum_\pm \left<V_{P\pm \frac{\beta}{2}}\right>_\sigma\ ,
\end{align}
where $\bar{\mu}_\sigma$ is the OPE coefficient
\begin{equation}
    V^d_{\langle 2,1\rangle} \to \bar\mu_\sigma \psi^d_{\langle1,1\rangle} + \tilde{\bar\mu}_\sigma \psi^d_{\langle3,1\rangle} \;.
\end{equation}
In bulk critical loop models, the absence of $V^d_{\langle 2,1\rangle}$ also leads to large spaces of solutions. But in the case of the 3-point function of diagonal fields $\left<V_{P_1}V_{P_2}V_{P_3}\right>^\text{sphere}$, we know that the correct solution obeys the shift equation coming from both $V^d_{\langle1,2\rangle}$ and $V^d_{\langle2,1\rangle}$, and therefore coincides with the 3-point function of $c\leq 1$ Liouville theory. This motivates
\begin{quote}
 \textbf{Main assumption: In the presence of a reflecting boundary,\\ disc 1-point functions obey two shift equations.}
\end{quote}
%\hs{I think this is related to the remarks around eq. (2.36) in the notes Boundary-Physics.tex, and the relationship between boundary conditions and topological defects. As for the double degeneracy, it must be related with the fact that $V^d_{<12>}$ is degenerate, while the $21$ defects also obey degenerate fusion equations. I need to think of this more.}
If $\bar\mu_\sigma = 2\cos(2\pi \sigma \beta)$, solutions of both shift equations are 
\begin{align}
\left<V_P\right>_\sigma = \sin(4\pi \sigma P) \;,
\end{align}
i.e. the infinite linear combination \eqref{sisz} reduces to one non-vanishing term. The only remaining ambiguities come from rescalings $\left<V_P\right>_\sigma\to N_\sigma\left<V_P\right>_\sigma$, where the factor $N_\sigma$ is $P$-independent because it must be invariant under $P\to P+\beta$ and $P\to P+\beta^{-1}$.

We can also compute disc 1-point functions of degenerate fields. Since $V^d_{\langle 1,1\rangle}$ and $\psi^d_{\langle 1,1\rangle}$ are respectively the bulk and boundary identity fields, the bulk-to-boundary coefficient $\mu_\sigma$ \eqref{eq:V12 to bndry} may be rewritten as
\begin{align}
 \mu_\sigma = \frac{\left<V_{\langle1,2\rangle}^d\right>_\sigma}{\left<\psi_{\langle 1,1\rangle}^d\right>_\sigma} = \frac{\left<V_{\langle1,2\rangle}^d\right>_\sigma}{\left<V_{\langle 1,1\rangle}^d\right>_\sigma} \ .
\label{mvv}
\end{align}
The first equality comes from using the OPE \eqref{eq:V12 to bndry} in $\left<V_{\langle1,2\rangle}^d\right>_\sigma$, taking into account $\left<\psi^d_{\langle 1,3\rangle}\right>_\sigma=0$ (since $\Delta_{(1,3)}\neq 0$). The second equality comes from taking the bulk identity field to the boundary and producing the boundary identity. In the case of 1-point functions of degenerate fields, the shift equation therefore reads
\begin{align}
 \frac{\left<V_{\langle1,2\rangle}^d\right>_\sigma}{\left<V_{\langle 1,1\rangle}^d\right>_\sigma} \left<V_{\langle1,s\rangle}^d\right>_\sigma = \left<V_{\langle1,s+1\rangle}^d\right>_\sigma + \left<V_{\langle1,s-1\rangle}^d\right>_\sigma\ ,
\end{align}
and the solution is
\begin{align}
 \left<V_{\langle1,s\rangle}^d\right>_\sigma= \frac{\sin(2\pi\beta^{-1}\sigma s)}{\sin(2\pi\beta^{-1}\sigma)}\left<V_{\langle 1,1\rangle}^d\right>_\sigma \ ,
\end{align}
where the disc partition function $\left<V_{\langle 1,1\rangle}^d\right>_\sigma$ is left unconstrained.

In section \ref{sec:modular} we derive boundary spectra using the annulus partition function. Here we will propose a technically simple criterion to distinguish between continuous and discrete reflecting boundaries, using disc 1-point functions.
We make the plausible but unproven assumption:
\begin{align}
 \text{The boundary } \sigma \text{ is discrete} \iff \lim_{P\to P_{(1,s)}} \left<V_P\right>_\sigma=\left<V^d_{\langle 1,s\rangle}\right>_\sigma \ .
 \label{ass}
\end{align}
If the boundary is discrete, we indeed expect that the bulk-to-boundary OPE commutes with the limit $P\to P_{(1,s)}$. If it is continuous, however, the bulk-to-boundary OPE involves an integral over boundary momenta, and poles of the integrand may well cross or pinch the integration contour in the limit $P\to P_{(1,s)}$, leading to more subtle behaviour.
Together with \eqref{mvv}, our assumption \eqref{ass} leads to the determination of the discrete values of $\sigma = P_{(0,S)}$ for $S\in\mathbb{N}^*$ in section \ref{sec:bndry fields}.

We can gain intuition about the behaviour of diagonal fields as they approach degenerate values by comparing with known results in Liouville theory. In Liouville theory with generic central charge $c\in \mathbb{C}\backslash (-\infty, 1]$, we have $\lim_{P\to P_{(r,s)}} V_P=V^d_{\langle r,s\rangle}$ in bulk OPEs; what about in disc 1-point functions? Disc 1-point functions obey the shift equation \eqref{mvsv}, in a field normalisation such that $V_P=V_{-P}$.
There exist discrete boundary conditions called Zamolodchikov--Zamolodchikov boundary conditions \cite{zz01}, parametrised by pairs of Kac indices $(m,n)\in (\mathbb{N}^*)^2$, and continuous boundary conditions called Fateev--Zamolodchikov--Zamolodchikov--Teschner boundary conditions \cite{fzz00, tes00}, parametrised by $\sigma\in\mathbb{C}$. Up to $P$-independent factors, the bulk 1-point functions and boundary cosmological constants are
\begin{align}
\text{ZZ:} & \ \left<V_P\right>_{(m,n)} = \sin(2\pi\beta m P)\sin(2\pi \beta^{-1}nP) \quad , \quad  \mu_{(m,n)}=2(-1)^m\cos(\pi \beta^{-2}n) \ ,
\label{zz}
\\
\text{FZZT:} & \ \left<V_P\right>_\sigma = \cos(4\pi \sigma P) \quad , \quad \mu_\sigma = 2\cos(2\pi \beta^{-1}\sigma)\ .
\label{fzzt}
\end{align}
In fact, a ZZ boundary condition is a linear combination of two FZZT boundary conditions with boundary parameters $\sigma = P_{(m,\pm n)}$. Since these two parameters correspond to the same boundary cosmological constant $\mu_{(m,n)}=\mu_{P_{(m,n)}}=\mu_{P_{(m,-n)}}$, the corresponding boundary conditions can be combined without spoiling the shift equations. The 1-point functions for ZZ and FZZT boundary conditions agree with our criterion \eqref{ass}: $\mu$ is a ratio of 1-point functions in the ZZ case but not in the FZZT case, except for $\sigma= P_{(0,s)}$ with $s\in\mathbb{Z}+\frac12$. And this exception would disappear if we also considered the shift equations from the second degenerate field $V^d_{\langle 2,1\rangle}$.

\subsection{Boundary spectra of annulus partition functions}\label{sec:modular}

On an annulus, consider a critical loop model with boundary conditions $\sigma_1$ on the inner boundary and $\sigma_2$ on the outer boundary:
\begin{align}
\begin{tikzpicture}[baseline=(current  bounding  box.center), thick,scale=1.2, every node/.style={scale=1}]
        \draw (0.8,0) arc [start angle=0,end angle=3600,x radius=0.8,y radius=0.8];
        \draw (2,0) arc [start angle=0,end angle=3600,x radius=2,y radius=2];
        \node at (-0.75,0.75) {$\sigma_1$};
        \node at (-1.6,1.6) {$\sigma_2$};
        \node at (0.9,.85) {$|z|{=}1$};
        \node at (2,1.75) {$|z| {=} e^{-i\pi\tau}$};
        \node at (-3.5,2) {$z$};
        \draw (-3.7,1.8) -- (-3.3,1.8) -- (-3.3,2.2);
\end{tikzpicture}
\end{align}
The corresponding annulus partition function can be computed in the bulk and boundary channels. The two channels are related by a modular transformation $\tau \rightarrow -\frac{1}{\tau}$, which will allow us to deduce the boundary spectrum from bulk structure constants.

We therefore start with the bulk channel. Since we are considering reflecting boundary conditions, only diagonal fields appear and we take the spectrum to be $\{V_P\}_{P\in P_0+\beta^{-1}\mathbb{Z}}$, as in the bulk CFT \cite{rib22}.
The partition function in the bulk channel is constructed from disc 1-point functions $\left<V_P\right>_\sigma$ and sphere 2-point functions $B_P$ as follows
\begin{align}
 Z_{\sigma_1, \sigma_2} = \sum_{P \in P_0 + \beta^{-1} \mathbb{Z}} \frac{\langle V_P \rangle_{\sigma_1} \langle V_P \rangle_{\sigma_2}}{B_P} \chi_P(\tau) \qquad \text{with} \qquad \chi_P(\tau) = \frac{e^{2\pi i \tau P^2}}{\eta(\tau)}\;,
 \label{zsisi}
\end{align}
where $\eta(\tau)$ is the Dedekind eta function. Using the expressions of the disc 1-point function \eqref{sol} and sphere 2-point structure constant \eqref{bp}, this is explicitly
\begin{align}
 Z_{\sigma_1, \sigma_2} =- \sin(\pi \beta^2)\sin(\pi \beta^{-2}) \sum_{P \in P_0 + \beta^{-1} \mathbb{Z}} \frac{\sin(4\pi \sigma_1 P) \sin(4\pi \sigma_2 P)}{\sin(2\pi \beta P) \sin(2\pi \beta^{-1}P)} \chi_P(\tau) \ .
 \label{zbulk}
\end{align}
In order to rewrite this partition function in the boundary channel, we use the behaviour of the character under the modular transformation:
\begin{equation}
    \chi_P(\tau) =  \sqrt{2}\int_{-\infty}^\infty dP'\; \cos(4\pi P P') \chi_{P'}(-\tfrac{1}{\tau})  \;.
\end{equation}
This leads to $Z_{\sigma_1,\sigma_2} = \int_{-\infty}^\infty dP'\; \rho_{\sigma_1,\sigma_2}(P')\chi_{P'}(-\tfrac{1}{\tau})$: an integral over a continuous boundary spectrum, with the spectral density
\begin{align}
\label{eq:rho}
 \rho_{\sigma_1,\sigma_2}(P')= -\frac{\sqrt{2} \sin(\pi \beta^2) \sin(\pi \beta^{-2})}{\sin(2\pi P_0 \beta)}\sum_{P\in P_0 + \beta^{-1} \mathbb{Z}} \frac{\sin(4\pi \sigma_1 P) \sin(4\pi \sigma_2 P)}{\sin(2\pi P \beta^{-1})}  \cos(4\pi P P')\; .
\end{align}
For generic $\sigma_i$ we cannot simplify the expression further. The boundary channel is an integral and therefore we have a continuous boundary spectrum. However, for special values of $\sigma$, the density is a combination of delta functions, and the integral collapses to a sum. This happens if and only if the summand in $\sum_{P\in P_0 + \beta^{-1} \mathbb{Z}}$ is in fact trigonometric, due to the identity
\begin{equation}\label{eq:delta}
  \sum_{P\in P_0 + \beta^{-1} \mathbb{Z}} e^{4\pi i PP'} = \frac{\beta}{2}  \sum_{r\in\mathbb{Z}} e^{-4\pi i P_0P_{(r,0)}} \delta\left(P'+P_{(r,0)}\right) \;.
\end{equation}
For the summand to be trigonometric, we need $\sigma_i = P_{(0,S_i)}$ with $S_i\in\mathbb{N}^*$ for $i=1$ or $i=2$. Then we can use the identity
\begin{equation}
    \frac{\sin(S_1 x) \sin(S_2 x)}{\sin(x)}
    \underset{\substack{S_1\in\mathbb{C}\\ S_2\in\mathbb{N}^*}}= \sum_{s \overset{2}{=} 1-S_2 }^{S_2 -1} \sin((S_1+s)x)
    \underset{S_1, S_2\in\mathbb{N}^*}= \sum_{s \overset{2}{=} |S_1-S_2| +1 }^{S_1 + S_2 -1} \sin(sx) \;.
\end{equation}
If the second boundary is discrete, the partition function becomes
\begin{align}
 Z_{\sigma_1, S_2}= -\frac{\beta \sin(\pi \beta^2) \sin(\pi \beta^{-2})}{\sqrt{2}\sin(2\pi P_0 \beta)}  \sum_{r\in\mathbb{Z}} \sum_{s \overset{2}{=} 1-S_2 }^{S_2 -1} \sin(2\pi r P_0 \beta) \chi_{\sigma_1+ P_{(r,s)}}(-\tfrac{1}{\tau}) \;.
\end{align}
If the first boundary is also discrete, the characters combine into characters of degenerate representations $\chi^d_{\langle r,s\rangle} = \chi_{P_{(r,s)}}-\chi_{P_{(r,-s)}}$, and we find
\begin{equation}
    Z_{S_1, S_2} = -\frac{\beta \sin(\pi \beta^2) \sin(\pi \beta^{-2})}{\sqrt{2}\sin(2\pi P_0 \beta)}  \sum_{r\in\mathbb{N^*}} \sum_{s\overset{2}{=} |S_1 - S_2| +1 }^{S_1 + S_2 -1}  \sin(2\pi r P_0 \beta) \chi^d_{\langle r,s\rangle}(-\tfrac{1}{\tau}) \;.
    \label{zss}
\end{equation}
We therefore recover the condition $\sigma = P_{(0,S)}$ with $S\in\mathbb{N}^*$ from section \ref{sec:bndry fields} for the boundary to be discrete. 

Moreover, in the case $S_1=S_2=S$, \eqref{zss} gives us the degenerate boundary spectrum 
\begin{align}
 \mathcal{S}_S = \left\{\psi^d_{\langle r,s\rangle}\right\}_{\substack{r\in \mathbb{N}^* \\ s = 1, 3, \dots, 2S-1}}\ .
\label{ss}
 \end{align}
This is the set of boundary fields given in table \eqref{tab:bndry fields}.

\subsection{Boundary spectra of disc 2-point functions}\label{sec:bou}

For a disc 2-point function $\left<V_{(r_1,s_1)}V_{(r_2,s_2)}\right>_{P_{(0,S)}}$ with a discrete reflecting boundary, the boundary channel decomposition \eqref{2pt-bdy} should involve a sum over the boundary spectrum $\mathcal{S}_S$ \eqref{ss}. However, the coefficients of some boundary fields may vanish, in which case the boundary channel spectrum is a subset $\mathcal{S}^{r_1,r_2}_S$ \eqref{s12} of $\mathcal{S}_S$.

The full spectrum $\mathcal{S}_S$ \eqref{ss} predicts bulk-to-boundary OPEs of the type
\begin{align}
 V_{(r_1,s_1)} \overset{S}{\to} \sum_{r\in\mathbb{N}^*}  \sum_{s \overset{2}{=} 1}^{2S-1} \psi^d_{\langle r,s\rangle}\ ,
\end{align}
where the $s$ sum increments in steps of $2$. But the boundary field $\psi^d_{\langle r,s\rangle}$ can contribute only if
\begin{equation}
    \langle V_{(r_1,s_1)} \psi^d_{\langle r,s\rangle} \rangle_{P_{(0,S)}} \neq 0 \;.
\end{equation}
Let us determine whether this is compatible with the fusion rules of the degenerate field $\psi^d_{\langle r,s\rangle}$.
Using the folding trick, this can be formulated in terms of a chiral 3-point function on the sphere. We have two chiral fields $\phi_{(\pm r_1,s_1)}$ with conformal dimensions $\Delta_{(\pm r_1,s_1)}$ (note that $\Delta_{(-r_1,s_1)} = \Delta_{(r_1,-s_1)}$) and a degenerate chiral field $\phi^d_{\langle r,s\rangle}$ with dimension $\Delta_{(r,s)}$. The 3-point function
\begin{equation}
    \langle \phi_{(r_1,s_1)} \phi_{(-r_1,s_1)} \phi^d_{\langle r,s\rangle} \rangle^\text{sphere} \;,
\end{equation}
is non-vanishing if and only if $\phi_{(-r_1,s_1)}$ appears in the OPE of $\phi^d_{\langle r,s\rangle}$ and $\phi_{(r_1,s_1)}$. The OPE is \cite{rib24}
\begin{equation}
    \phi^d_{\langle r,s\rangle} \phi_{(r_1,s_1)} = \sum_{r' \overset{2}{=} r_1 - r + 1}^{r_1 + r - 1} \sum_{s' \overset{2}{=} s_1 - s + 1}^{s_1 + s - 1} \phi_{(r',s')} \;.
\end{equation}
Therefore, for $r_1\geq 0$ we have $\phi_{(-r_1,s_1)} \in \phi^d_{\langle r,s\rangle} \phi_{(r_1,s_1)}$ if and only if $r \geq 2r_1 +1$ and $r - 2r_1$ is odd. Returning to bulk-to-boundary OPEs, this amounts to
\begin{align}
 V_{(r_1,s_1)} \overset{S}{\to} \sum_{r\overset{2}{=} 2r_1+1}^\infty  \sum_{s \overset{2}{=} 1}^{2S-1} \psi^d_{\langle r,s\rangle}\ .
 \label{b2b}
\end{align}
For the disc 2-point function $\left<V_{(r_1,s_1)}V_{(r_2,s_2)}\right>_{P_{(0,S)}}$, this OPE leads to the spectrum \eqref{s12}.

The OPE \eqref{b2b} is consistent with the geometrical interpretation of the bulk and boundary fields. If we have a bulk field $V_{(r_1,s_1)}$ emitting $2r_1$ legs, then as it approaches the boundary it will trap $r' \in \mathbb{N}$ curves. The result is $2r'$ more arcs attached to the boundary point, as shown in the figure.
\begin{align}
\begin{tikzpicture}[baseline=(current  bounding  box.center), scale = 1.4]
\draw[thick, red] plot [smooth] coordinates {(0, 0)(.2, .7)(.4, 1.2)};
\node at (.5, 1.3) {$\scriptstyle{1}$};
\draw[thick, red] plot [smooth] coordinates {(0, 0)(.3, .6)(.6, 1)};
\node at (.7, 1.1) {$\scriptstyle{2}$};
\draw[thick, red] plot [smooth] coordinates {(0, 0)(.4, .5)(.8, .8)};
\node at (.9, .9) {$\scriptstyle{3}$};
\draw[thick, red] plot [smooth] coordinates {(0, 0)(.5, .1)(1.1, .3)};
\node at (1.3, .3) {$\scriptstyle{2r_1}$};
 \draw (0, 0) node[fill, circle, minimum size = 1.6mm, inner sep = 0]{};
 \node[left] at (0, 0) {$V_{(r_1,s_1)}$};
 \draw[dashed] plot [smooth] coordinates {(.6, .6)(.7, .2)};
 \draw[thick,->] (0,0) -- (0,-0.25);
\draw[thick, red] plot [smooth] coordinates {(-1.5, -.2)(.2, -0.3)(1.4, -0.2)};
\node at (1.5, -.15) {$\scriptstyle{1}$};
\draw[dashed] (0,-0.6) -- (0,-0.35);
\draw[thick, red] plot [smooth] coordinates {(-1.5, -.6)(-.1, -0.7)(1.4, -0.6)};
\node at (1.55, -.55) {$\scriptstyle{r'}$};
\draw (-1.6, -0.8) -- (1.6, -0.8);
\end{tikzpicture}
\hspace{1cm}
\rightarrow
\hspace{1cm}
\begin{tikzpicture}[baseline=(current  bounding  box.center), scale = 1.4]
 \node[below] at (0, -.1) {$\psi^d_{\langle 2r_1+2r'+1,s'\rangle}$};
 \begin{scope}[rotate = 90]
 \draw[thick, red] plot [smooth] coordinates {(0, 0)(.2, .7)(.4, 1.2)};
\node at (.5, 1.3) {$\scriptstyle{1}$};
\draw[thick, red] plot [smooth] coordinates {(0, 0)(.3, .6)(.6, 1)};
\node at (.7, 1.1) {$\scriptstyle{2}$};
\draw[thick, red] plot [smooth] coordinates {(0, 0)(.4, .5)(.8, .8)};
\node at (.9, .9) {$\scriptstyle{3}$};
\draw[thick, red] plot [smooth] coordinates {(0, 0)(.5, -.3)(1.1, -.5)};
\node at (1.3, -.5) {$\scriptstyle{2r_1+2r'}$};
\draw[dashed] plot [smooth] coordinates {(.6, .45)(.75, .1)(.7, -.25)};
\end{scope}
 \draw (-1.6, 0) -- (1.6, 0);
 \draw (0, 0) node[fill, circle, minimum size = 1.6mm, inner sep = 0]{};
\end{tikzpicture}
\end{align}
Hence, as it approaches the boundary it becomes an operator that emits $2r_1 + 2r'$ arcs. This operator is the degenerate boundary field $\psi^d_{\langle2r_1+2r'+1,s'\rangle}$. This geometrical argument gives no constraint on $s'$. In the loop model we sum over all curve configurations and hence over all $r'\geq0$.

\section{Numerical checks}\label{sec:num checks}

For a disc 2-point function with discrete reflecting boundaries, we will now numerically check crossing symmetry, i.e. the agreement between the bulk and boundary channel expansions \eqref{2pt-bulk} and \eqref{2pt-bdy}. Let us first outline the relevant numerical techniques.

\subsection{Numerical method}

Numerical bootstrap methods for sphere four-point functions in 2d CFTs with exactly known spectra are reviewed in \cite{rib24}. We have adapted these methods to disc 2-point functions. We assume that the bulk and boundary spectra are known and discrete, as we would not know how to safely truncate a continuous spectrum to make it finite: this is why we are focusing on discrete boundaries.

Schematically we have the following. In the bulk channel
\begin{equation}
    \langle V_{(r_1,s_1)} V_{(r_2,s_2)} \rangle_\sigma = \sum_{k \in \mathcal{S}_{\text{bulk}}} D^{(s)}_k \mathcal{F}^{(s)}_k(z) \;,
\end{equation}
where $\mathcal{S}_{\text{bulk}}$ is the bulk channel spectrum, $D^{(s)}_k$ are bulk channel four-point structure constants and $\mathcal{F}^{(s)}_k(z)$ are s-channel conformal blocks (previously denoted by diagrams \begin{tikzpicture}[baseline=(base), scale = .16]
\coordinate (base) at (0, -0.5);
\draw (-1,2) -- (0,0) -- (4, 0) -- (5,2);
\draw (-1,-2) -- (0,0);
\draw (4,0) -- (5,-2);
 \end{tikzpicture} in \eqref{2pt-bulk}). $z$ is the cross ratio
 \begin{equation}
     z=\left|\frac{z_1-z_2}{z_1-\bar z_2}\right|^2 \in (0, 1)
 \end{equation}
 where $z_1$ and $z_2$ are the positions of the two fields. We have suppressed the dependence of $\mathcal{F}^{(s)}_k$ on the fields in the 2-point function.

In the boundary channel
\begin{equation}
    \langle V_{(r_1,s_1)} V_{(r_2,s_2)} \rangle_\sigma = \sum_{k \in \mathcal{S}_{\text{bdy}}} D^{(t)}_k \mathcal{F}^{(t)}_k(z) \;,
\end{equation}
where $\mathcal{S}_{\text{bdy}}$ is the boundary channel spectrum, $D^{(t)}_k$ are boundary channel four-point structure constants and $\mathcal{F}^{(t)}_k(z)$ are t-channel conformal blocks (previously denoted by diagrams \begin{tikzpicture}[baseline=(base), scale = .1]
\coordinate (base) at (0, 1.4);
\draw (2, -1) -- (0,0) -- (0, 4) -- (2,5);
\draw (-2, -1) -- (0,0);
\draw (0,4) -- (-2,5);
 \end{tikzpicture} in \eqref{2pt-bdy}).
Crossing symmetry is the equation
\begin{equation}
    \sum_{k \in \mathcal{S}_{\text{bulk}}} D^{(s)}_k \mathcal{F}^{(s)}_k(z) = \sum_{k \in \mathcal{S}_{\text{bdy}}} D^{(t)}_k \mathcal{F}^{(t)}_k(z) \;.
\end{equation}
The spectra and conformal blocks are considered to be known and the structure constants are unknown. We truncate both spectra and evaluate the crossing equation at finitely many values of $z$.
This allows us to numerically determine the dimension of the space of solutions, and the structure constants for specific solutions. In practice we focus on cases where the solution is unique, by fixing the values of some structure constants and/or by restricting the boundary spectrum.

For a disc 2-point function, the conformal blocks should in principle be evaluated for values of the cross-ratio $z=\left|\frac{z_1-z_2}{z_1-\bar z_2}\right|^2 \in (0, 1)$. By analyticity, crossing symmetry in fact holds for any $z\in\mathbb{C}\backslash \{0, 1\}$, and we find it numerically convenient to use complex values of $z$.

\subsection{Numerical claims}

We structure our numerical checks of crossing symmetry into 3 claims. We then make a final claim, claim \ref{claim4}, based on numerical observations that were made while verifying claims \ref{claim1} -- \ref{claim3}.

\begin{theorem}\label{claim1}
\textbf{Given the bulk channel decomposition \eqref{2pt-bulk} and boundary spectrum $\mathcal{S}^{r_1,r_2}_S$ \eqref{s12}, there is a unique solution to the crossing equation \eqref{2pt-bulk} = \eqref{2pt-bdy}.}
\end{theorem}
Here the bulk channel expansion is fixed to be that of \eqref{2pt-bulk}. The bulk spectrum is known and the structure constants in front of the conformal blocks are known. In the boundary channel we fix the boundary spectrum to be \eqref{s12}. The structure constants $D_k$ in \eqref{2pt-bdy} are unknowns that we solve for numerically.

This verifies that our explicit expression \eqref{2pt-bulk} gives a consistent CFT correlator and hence we have an explicit expression for the disc 2-point function as claimed in section \ref{sec:2pnt function}.

We tested this claim with parameters $\beta = 0.8 +0.1i$, $S\in \{1, 2,3\}$ and bulk channel momentum $P_0=P_{(0, \frac{1}{17})} $, for the following disc 2-point functions:
\begin{align}
&
\left\langle  V_{P_{(0, \frac{1}{15})}} V_{P_{(0, \frac{1}{16})}} \right\rangle
\; , \;
\left\langle  V_{P_{(0, \frac{1}{15})}} V_{(1,0)} \right\rangle
\; , \;
\left\langle  V_{(1, 0)} V_{(1,0)} \right\rangle
\; , \;
\left\langle  V_{(\frac12 , 0)} V_{(\frac32, \frac23)} \right\rangle
\; , \;
\left\langle  V_{(\frac12 , 2)} V_{(\frac12, 2)} \right\rangle
\ , \nonumber
\\
&
\left\langle  V_{P_{(0, \frac{1}{15})}} V_{(2, 0)} \right\rangle
\; , \;
\left\langle  V_{(2, \frac12)} V_{(2, -\frac12)} \right\rangle
\; , \;
\left\langle  V_{(3, \frac23)} V_{(3, \frac13)} \right\rangle
\; , \;
\left\langle  V_{(1, 1)} V_{(3, \frac13)} \right\rangle
\; , \;
\left\langle  V_{(2, \frac12)} V_{(3, \frac23)} \right\rangle
\ , \nonumber
\\
&
\left\langle  V_{(\frac12 , 0)} V_{(\frac12, 0)} \right\rangle
\; , \;
\left\langle  V_{(\frac12 , 0)} V_{(\frac32, 0)} \right\rangle
\; , \;
\left\langle  V_{(\frac12 , 0)} V_{(\frac52, 0)} \right\rangle
\; , \;
\left\langle  V_{(\frac32 , 0)} V_{(\frac32, 0)} \right\rangle
\; , \;
\left\langle  V_{(\frac32 , 0)} V_{(\frac52, 0)} \right\rangle
\ .
\label{list2pt0}
\end{align}

\begin{theorem}\label{claim2}
\textbf{Given the bulk channel decomposition \eqref{2pt-bulk} and boundary spectrum $\mathcal{S}_S$ \eqref{ss}, there is a unique solution to the crossing equation. For this solution, $D_{\left<r,s\right>}=0$ if $r\leq 2\max(r_1,r_2)$ or if $r\equiv 2 r_1\bmod 2$. Therefore the boundary spectrum is $\mathcal{S}_S^{r_1,r_2}\subset\mathcal{S}_S$ \eqref{s12}.}
\end{theorem}
Here we have tested whether extending the spectrum to the full set of boundary fields that can live on a discrete reflecting boundary gives new solutions to the crossing equations. We find that this is not the case and therefore no solutions are lost when we restrict to the boundary spectrum $S^{r_1,r_2}_S$ in the boundary channel of the correlator \eqref{2pt-bdy}.

To give a quantitative idea of our evidence for this claim, we focus on the non-vanishing boundary structure constant with the lowest Kac indices $D_\text{min} = D_{\langle 2\max(r_1,r_2)+1,1\rangle}$. For this structure constant we display the deviation: an estimate of the relative error of the numerical bootstrap result with respect to the true value \cite{rib24}. If claim \ref{claim2} is true we expect small deviations; if it were wrong the deviations would typically be no smaller than about $10^{-3}$.
Our numerical calculations use numbers with $100$ decimal digits, and the spectrum is truncated to a maximum conformal dimension of $80$.
\begin{align}\renewcommand{\arraystretch}{2.0}
  \begin{tabular}{|c|c|c|c|c|}
    \hline
    \multirow{2}{*}{2-point function} &
          \multirow{2}{*}{$D_{\text{min}}$}&
      \multicolumn{3}{c|}{Deviation}
       \\
       \cline{3-5}
       &
	&
       $S=1$
       &
        $S=2$
        &
        $S=3$
       \\ \hline\hline
       $  \left\langle V_{(\frac12, 2)}V_{(\frac12, 2)} \right\rangle$
  &  
    $D_{\left\langle 2,1 \right\rangle}$
   &   
$   10^{-29}$
   &
$   10^{-19}$
   &
$   10^{-6}$
   \\
$  \left\langle V_{(\frac12, 2)}V_{(\frac32, \frac23)} \right\rangle$
  &  
    $D_{\left\langle 4,1 \right\rangle}$
   & 
   $ 10^{-34}$
   &
   $ 10^{-25}$
   &
   $ 10^{-10}$
     \\
$  \left\langle V_{(\frac32, \frac23)}V_{(\frac32, \frac23)} \right\rangle$
  & 
    $D_{\left\langle 4,1 \right\rangle}$
    &  
       $ 10^{-37}$
    &
       $ 10^{-26}$
    &
       $ 10^{-8}$
     \\
$  \left\langle V_{(1, 0)}V_{(1, 0)} \right\rangle$
  &  
    $D_{\left\langle 3,1 \right\rangle}$
   & 
        $ 10^{-45}$
   &
        $ 10^{-32}$
   &
        $ 10^{-11}$
     \\
$  \left\langle V_{(1, 0)}V_{(2, 0)} \right\rangle$
  &  
    $D_{\left\langle 5,1 \right\rangle}$
   &  
        $ 10^{-43}$
   &
        $ 10^{-31}$
   &
        $ 10^{-13}$
    \\
$  \left\langle V_{(2, 0)}V_{(2, 0)} \right\rangle$
  & 
    $D_{\left\langle 5,1 \right\rangle}$
    &
         $ 10^{-41}$
    &
         $ 10^{-31}$
    &
         $ 10^{-13}$
       \\
    \hline
  \end{tabular}
\end{align}
In some examples, we find more structure constants that vanish, and the boundary spectrum becomes even smaller. In the cases of $\left<V_{(\frac12,0)}V_{(\frac12,0)}\right>, \left<V_{(\frac12,0)}V_{(\frac32,0)}\right>$ and $\left<V_{(\frac32,0)}V_{(\frac32,0)}\right>$ with $S=2$ we find $D_{\langle r, 1\rangle}=0$ and, when $S = 3$, we find $D_{\langle r, 3\rangle}=0$. Understanding this observation is left for future work.

\begin{theorem}\label{claim3}
    \textbf{In the case of $\left<V_{P_1}V_{P_2}\right>_{P_{(0,S)}}$, the value of $D_{\langle 1,1\rangle}$ is given by Eq. \eqref{d11}.}
\end{theorem}
When both external fields are diagonal ($r_1 = r_2 = 0$), the degenerate boundary field $\psi^d_{\langle1,1\rangle}$ is in the boundary spectrum \eqref{s12}. In this case, we can compute the structure constant corresponding to $\psi^d_{\langle 1,1\rangle}$ in terms of the disc 1-point function:
\begin{equation}\begin{split}
 D_{\langle 1,1\rangle} &= \frac{\left<V_{P_1}\psi^d_{\langle 1,1\rangle}\right>_{P_{(0,S)}} \left<V_{P_2}\psi^d_{\langle 1,1\rangle}\right>_{P_{(0,S)}}}{\left<\psi^d_{\langle 1,1\rangle}\psi^d_{\langle 1,1\rangle}\right>_{P_{(0,S)}}} =
 \frac{\left<V_{P_1}\right>_{P_{(0,S)}} \left<V_{P_2}\right>_{P_{(0,S)}}}{\left<V^d_{\langle 1,1\rangle}\right>_{P_{(0,S)}}}\\
 &=
 \frac{\left<V_{P_1}\right>_{P_{(0,S)}} \left<V_{P_2}\right>_{P_{(0,S)}}}{\left<V_{P_{(1,1)}}\right>_{P_{(0,S)}}} \;.
\end{split}\end{equation}
The first equality comes from taking the bulk fields to the boundary and producing the identity field. In the second equality we used the fact that $\psi^d_{\langle 1,1\rangle}$ and $V^d_{\langle 1,1\rangle}$ are identity fields and in the third equality the characterisation \eqref{ass} of discrete boundaries was used to swap the identity field for a diagonal field with momentum $P_{(1,1)}$. (We discussed the relation between diagonal and degenerate fields in section \ref{sec:dopf}.) With our disc 1-point function \eqref{sol}, this leads to \eqref{d11}.

In a few cases, let us display the error, i.e. the relative difference between our analytic prediction and the numerical result for $D_{\langle 1,1\rangle}$. Claim \ref{claim3} implies that the errors should be small, and should decrease when the numerical precision is improved. Numerical results are consistent with these expectations.

\begin{align}\renewcommand{\arraystretch}{2.0}
  \begin{tabular}{|c|c|c|c|}
    \hline
    \multirow{2}{*}{2-point function} &
      \multicolumn{3}{c|}{Error} 
       \\
       \cline{2-4}
       &
       $S=1$
       &
        $S=2$
        &
        $S=3$
       \\ \hline
       $  \left\langle V_{P_{(0, \frac{1}{12})}}V_{P_{(0, \frac{1}{13})}} \right\rangle$
  &  
$  10^{-45}$
   & 
        $ 10^{-30}$
   &
       $ 10^{-10}$
     \\
$  \left\langle V_{P_{(0, \frac{1}{22})}}V_{P_{(0, \frac{1}{23})} }\right\rangle$
  &  
       $ 10^{-44}$
   &
        $ 10^{-30}$
   &
        $ 10^{-9}$
      \\
$  \left\langle V_{P_{(0, \frac{1}{9})}}V_{P_{(0, \frac{1}{17})}} \right\rangle$
  &  
       $ 10^{-45}$
   &
        $ 10^{-30}$
   &
        $ 10^{-10}$
      \\
\hline
\end{tabular}
 \label{Cstu_wx}
\end{align}

\begin{theorem}\label{claim4}

\textbf{The ratios of boundary structure constants $\frac{D_{\langle r,s\rangle}}{D_{\langle r',s\rangle}}$, for a given correlator, do not depend on the boundary parameter $S$.}
\end{theorem}
We have tested our claim in the following cases:
\begin{align}
&
\left\langle  V_{(1, 1)} V_{(1, 1)} \right\rangle
\; , \;
\left\langle  V_{(1, 1)} V_{(3, \frac13)} \right\rangle
\; , \;
\left\langle  V_{(2, 1)} V_{(2, \frac12)} \right\rangle
\; , \;
\left\langle  V_{(2, 0)} V_{(3, 0)} \right\rangle
\ , \nonumber \\
&
\left\langle   V_{P_{(0, \frac15)}} V_{(1, 1)}\right\rangle
\; , \;
\left\langle  V_{P_{(0, \frac17)}} V_{(3, \frac13)} \right\rangle
\; , \;
\left\langle  V_{P_{(0, \frac18)}} V_{(2, 0)} \right\rangle
\; , \;
\left\langle  V_{P_{(0, \frac19)}} V_{(2, \frac12)} \right\rangle
\ .\nonumber
\label{list2pt1}
\end{align}
We will give some more detailed numerical results for a specific subcase of our claim. Our claim implies that the following quantity should vanish:
\begin{align}
 \text{Error} = \left|1- \frac{D_{\langle r',1\rangle}^{(S)}}{D_{\langle r,1\rangle}^{(S)}} \frac{D_{\langle r,1\rangle}^{(S=1)}}{D_{\langle r',1\rangle}^{(S=1)}} \right|\ .
\end{align}
We now display the order of magnitude of the error for a couple of 2-point functions. We find errors that are close to zero, supporting our claim:
\begin{align}
\renewcommand{\arraystretch}{2.2}
\begin{array}{cc}
\left\langle  V_{P_{(0, \frac{1}{15})}} V_{P_{(0, \frac{1}{16})}} \right\rangle
\hspace{0.5cm}
&
\left\langle  V_{(1, 0)} V_{(1, 0)} \right\rangle
\vspace{0.5cm}
\\
  \begin{tabular}{|c|c|c|}
    \hline
    \multirow{2}{*}{$(r,s)$} &
      \multicolumn{2}{c|}{Error}
       \\
       \cline{2-3}
       &
       $S=2$
       &
        $S=3$
       \\ 
       \hline
 $  (3, 1)$
  &  
  $10^{-31}$
   &
  $10^{-10}$
      \\
$ (5, 1)$
  &  
  $10^{-30}$
   &
     $10^{-9}$
      \\
$ (7, 1)$
  &  
    $10^{-29}$
   &
     $10^{-8}$
      \\
$ (9, 1)$
  &  
    $10^{-26}$
   &
     $10^{-6}$
      \\
\hline
\end{tabular}
&
  \begin{tabular}{|c|c|c|}
    \hline
    \multirow{2}{*}{$(r,s)$} &
      \multicolumn{2}{c|}{Error}
       \\
       \cline{2-3}
       &
       $S=2$
       &
        $S=3$
       \\ 
       \hline
 $  (5, 1)$
  &  
       $10^{-28}$
   &
        $10^{-8}$
      \\
$ (7, 1)$
  &  
       $10^{-27}$
   &
        $10^{-7}$
      \\
$ (9, 1)$
  &  
       $10^{-24}$
   &
        $10^{-6}$
      \\
$ (11, 1)$
  &  
       $10^{-21}$
   &
        $10^{-5}$
      \\
\hline
\end{tabular}
\end{array}
\label{tablin}
\end{align}
As a consequence of this claim, for any two bulk fields $V_1,V_2$ and $S\in \mathbb{N}^*$, there exist coefficients $c_1,c_2,\dots, c_{S}$ such that the linear combination
$
 \sum_{S'=1}^{S} c_{S'} \left<V_1V_2\right>_{P_{(0,S')}}
$
has the boundary spectrum $\{\psi^d_{\langle r,2S-1\rangle}\}_{r\in 2\max(r_1,r_2)+1+2\mathbb{N}}$, which is the subset of $\mathcal{S}^{r_1,r_2}_S$ \eqref{s12} where the second Kac index is fixed to its largest value. If we indeed determine $c_{S'}$ such that $\psi^d_{\langle r_0,s<2S-1\rangle}$ disappear from the spectrum for some $r_0$, then our claim ensures that all $\psi^d_{\langle r,s<2S-1\rangle}$ disappear as well.

\section{Boundary conditions in lattice loop models}
\label{sec:latt}

While we have borrowed terminology and used some intuition from loop models, our analysis was done purely in the bootstrap formalism. As usual with this axiomatic formalism, a separate analysis is needed for identifying the physical systems that our results describe.
We will now propose a construction of our reflecting boundary conditions in lattice loop models. We will then review the boundary conditions that are already known in these models, and determine whether they are reflecting.

To compare CFT and lattice results, we will focus on the bulk channel decomposition of the annulus partition function:
\begin{itemize}
 \item If both boundaries of the annulus are reflecting, this takes the form
$Z_{\sigma_1,\sigma_2}$ \eqref{zbulk}, which is a sum over momenta $P\in P_0+\beta^{-1}\mathbb{Z}$ that differ by multiples of $\beta^{-1}$.
\item If both boundaries are sticky, we have terms with $P=P_{(r,0)}=\frac12 r\beta$ for some $r\in\frac12\mathbb{N}^*$.
\end{itemize}
(The value of $\beta$ is deduced from the loop weight $n$ \eqref{nwp}.)

\subsection{Lattice construction of discrete reflecting boundaries}

The simplest boundary in lattice loop models is called a free boundary. Such a boundary does not change any property of the model, except the geometry on which it is defined. As we will see in Section \ref{sec:ckbc}, a free boundary corresponds to our discrete reflecting boundary with parameter $S=1$. To build other discrete reflecting boundaries, we have to modify interactions near the boundary.
We will describe these modifications in the transfer matrix formalism, assuming we are in the dense phase $0 < \beta^2<1$.

\subsubsection*{Modified transfer matrix}

Let us build a lattice with 2 boundaries by considering the time evolution of an open spin chain with $2N$ sites. The loop model's evolution is described by a transfer matrix, written in terms of the generators $1, e_1, e_2, \ldots, e_{2N-1}$ of the Temperley--Lieb algebra $\text{TL}_{2N}(n)$. With free boundaries, the transfer matrix is
\begin{equation}
    T_1 = \prod_{i\overset{2}{=}2}^{2N-2} (1+e_i) \prod_{i\overset{2}{=}1}^{2N-1} (1+e_i) \;.
\end{equation}
For a discrete reflecting boundary with parameter $S\geq 2$ on the left of the chain, we propose that the factors with $1\leq i\leq S-1$ should be replaced by the Jones--Wenzl projector $P_S$ \cite{morrison2015}. This leads to the transfer matrix
\begin{align}
 T_S = P_S \prod_{i\overset{2}{=}2\lfloor\frac{S+1}{2}\rfloor}^{2N-2} (1+e_i) \prod_{i\overset{2}{=}1+ 2\lfloor \frac{S}{2} \rfloor}^{2N-1} (1+e_i) \ ,
\end{align}
where $P_2 = 1-\frac{1}{n} e_1$, and more generally $P_S$ is the unique element of $\text{TL}_S(n)$ satisfying $P_S^2=P_S$ and $e_iP_S=P_Se_i=0$ for $i=1,2,\ldots,S-1$.
(In particular $P_S$ is a central idempotent.)
For two reflecting boundaries with parameters $S_1$ on the left and $S_2$ on the right, we define
\begin{align}
 \label{TS1S2}
 T_{S_1,S_2} = P^\text{left}_{S_1} P^\text{right}_{S_2} \prod_{i\overset{2}{=}2\lfloor\frac{S_1+1}{2}\rfloor}^{2N-2\lfloor\frac{S_2+1}{2}\rfloor} (1+e_i) \prod_{i\overset{2}{=}1+ 2\lfloor \frac{S_1}{2} \rfloor}^{2N-1-2\lfloor\frac{S_2}{2}\rfloor} (1+e_i) \ ,
\end{align}
where $P^\text{left}_{S_1}$ projects onto the first $S_1$ sites as before, whereas $P^\text{right}_{S_2}$ now projects onto the last $S_2$ sites.

Our construction generalises the so-called new boundary condition of the $3$-state Potts model \cite{Ian_Affleck_1998}, which corresponds to the case $S=2$.

\subsubsection*{Boltzmann weight interpretation}

Let us interpret the transfer matrix $T_{S,S}$ \eqref{TS1S2} in terms of microscopic Boltzmann weights. This applies to the loop model, and also to the  $Q$-state Potts model with $Q=n^2$, which is formulated in terms of Fortuin--Kasteleyn clusters. The loop model's transfer matrix generates a diagonally oriented square lattice. This is the medial lattice of an axially oriented square lattice, whose vertices are the Potts spins. Each lattice edge
between two neighbouring Potts spins (shown as black circles) carries the interaction
\begin{equation}
 \label{bulk-edge}
 \check{R}_i = 1 + e_i  =
 \begin{tikzpicture}[scale=1.0,baseline=-0.6ex]
 \fill (0,0) circle (0.1cm);
 \fill (1,0) circle (0.1cm);
 \draw[thick, red] (0.0,0.5) -- (0.3,0.2) arc [start angle=45, end angle=-45, radius=0.3] -- (0.0,-0.5);
 \draw[thick, red] (1.0,0.5) -- (0.7,0.2) arc [start angle=135, end angle=225, radius=0.3] -- (1.0,-0.5);
 \end{tikzpicture} +
 \begin{tikzpicture}[scale=1.0,baseline=-0.6ex]
 \draw[ultra thick,blue] (0,0)--(1,0);
 \fill (0,0) circle (0.1cm);
 \fill (1,0) circle (0.1cm);
 \draw[thick, red] (0.0,0.5) -- (0.3,0.2) arc [start angle=-135, end angle=-45, radius=0.3] -- (1.0,0.5);
 \draw[thick, red] (0.0,-0.5) -- (0.3,-0.2) arc [start angle=135, end angle=45, radius=0.3] -- (1.0,-0.5);
\end{tikzpicture}
\quad \mbox{ or } \quad
 \begin{tikzpicture}[scale=1.0,baseline=-0.6ex]
 \fill (0.5,-0.5) circle (0.1cm);
 \fill (0.5,0.5) circle (0.1cm);
 \draw[ultra thick,blue] (0.5,-0.5)--(0.5,0.5);
 \draw[thick, red] (0.0,0.5) -- (0.3,0.2) arc [start angle=45, end angle=-45, radius=0.3] -- (0.0,-0.5);
 \draw[thick, red] (1.0,0.5) -- (0.7,0.2) arc [start angle=135, end angle=225, radius=0.3] -- (1.0,-0.5);
 \end{tikzpicture} +
 \begin{tikzpicture}[scale=1.0,baseline=-0.6ex]
 \fill (0.5,-0.5) circle (0.1cm);
 \fill (0.5,0.5) circle (0.1cm);
 \draw[thick, red] (0.0,0.5) -- (0.3,0.2) arc [start angle=-135, end angle=-45, radius=0.3] -- (1.0,0.5);
 \draw[thick, red] (0.0,-0.5) -- (0.3,-0.2) arc [start angle=135, end angle=45, radius=0.3] -- (1.0,-0.5);
\end{tikzpicture}
\end{equation}
where the first picture is for a horizontal edge (odd $i$) and the second one for a vertical edge (even $i$).
Fortuin--Kasteleyn clusters are built from the blue edges, while Temperley--Lieb
loops are built from the red curves.
The figure \eqref{bulk-edge} shows that there is a one-to-one correspondence between configurations of FK clusters and TL loops. This immediately implies that the corresponding lattice algebras are isomorphic, so below we can freely switch between the two formulations, provided we make sure that the respective Boltzmann weights are equivalent.

Let us draw the lattices on which the loops and clusters live, in the cases $S=1,2,3,4$.
In the cases $S=2,3,4$ with non-free boundaries, the weights are not simply given by \eqref{bulk-edge}: in addition, there are grey areas that denote multi-spin interactions, and come with their own weights. The grey areas are polygons of degree $S$, whose $S$ vertices interact as dictated by the Jones--Wenzl projector $P_S$ from the transfer matrix $T_{S,S}$ \eqref{TS1S2}:
\begin{align}
\begin{tikzpicture}[scale=0.95]
 \foreach \x in {0,1,2,3,4}
  \draw (\x,0) -- (\x,3);
 \foreach \y in {0,1,2,3}
  \draw (0,\y) -- (4,\y);
 \foreach \x in {0.5,1.5,2.5,3.5}
 \foreach \y in {0,1,2,3}
 {
  \draw[thick,red] (\x-0.25,\y-0.25) -- (\x+0.25,\y+0.25);
  \draw[thick,red] (\x+0.25,\y-0.25) -- (\x-0.25,\y+0.25);
 }
 \foreach \x in {0,1,2,3,4}
 \foreach \y in {0.5,1.5,2.5}
 {
  \draw[thick,red] (\x-0.25,\y-0.25) -- (\x+0.25,\y+0.25);
  \draw[thick,red] (\x+0.25,\y-0.25) -- (\x-0.25,\y+0.25);
 }
 \foreach \y in {0,1,2,3}
 {
  \draw[thick, red, rounded corners=0.2cm] (-0.25,\y-0.25) -- (-0.5,\y) -- (-0.25,\y+0.25);
  \draw[thick, red, rounded corners=0.2cm] (4.25,\y-0.25) -- (4.5,\y) -- (4.25,\y+0.25);
 }
 \node at (-0.25,-0.5) {\tiny $1$};
 \node at (0.25,-0.5) {\tiny $2$};
 \node at (0.75,-0.5) {\tiny $3$};
 \node at (1.25,-0.5) {\tiny $4$};
 \node at (1.75,-0.5) {\tiny $5$};
 \node at (3,-0.5) {\tiny $\cdots$};
 \node at (4.25,-0.5) {\tiny $2N$};
 \foreach \x in {0,1,2,3,4}
 \foreach \y in {0,1,2,3}
 {
  \fill (\x,\y) circle (0.1cm);
 }
\end{tikzpicture}
& \qquad
\begin{tikzpicture}[scale=0.95]
 \foreach \x in {1,2,3}
  \draw (\x,0) -- (\x,3);
 \foreach \y in {0,1,2,3}
  \draw (0,\y) -- (4,\y);
 \foreach \y in {0,1,2}
 {
  \draw[fill=black!10] (0,\y+0.5) ellipse (0.1cm and 0.5cm); 
  \draw[fill=black!10] (4,\y+0.5) ellipse (0.1cm and 0.5cm); 
 }
 \foreach \x in {0.5,1.5,2.5,3.5}
 \foreach \y in {0,1,2,3}
 {
  \draw[thick,red] (\x-0.25,\y-0.25) -- (\x+0.25,\y+0.25);
  \draw[thick,red] (\x+0.25,\y-0.25) -- (\x-0.25,\y+0.25);
 }
 \foreach \x in {0,1,2,3,4}
 \foreach \y in {0.5,1.5,2.5}
 {
  \draw[thick,red] (\x-0.25,\y-0.25) -- (\x+0.25,\y+0.25);
  \draw[thick,red] (\x+0.25,\y-0.25) -- (\x-0.25,\y+0.25);
 }
 \foreach \y in {0,1,2,3}
 {
  \draw[thick, red, rounded corners=0.2cm] (-0.25,\y-0.25) -- (-0.5,\y) -- (-0.25,\y+0.25);
  \draw[thick, red, rounded corners=0.2cm] (4.25,\y-0.25) -- (4.5,\y) -- (4.25,\y+0.25);
 }
 \node at (-0.25,-0.5) {\tiny $1$};
 \node at (0.25,-0.5) {\tiny $2$};
 \node at (0.75,-0.5) {\tiny $3$};
 \node at (1.25,-0.5) {\tiny $4$};
 \node at (1.75,-0.5) {\tiny $5$};
 \node at (3,-0.5) {\tiny $\cdots$};
 \node at (4.25,-0.5) {\tiny $2N$};
 \foreach \x in {0,1,2,3,4}
 \foreach \y in {0,1,2,3}
 {
  \fill (\x,\y) circle (0.1cm);
 }
\end{tikzpicture}
\\
\begin{tikzpicture}[scale=0.95]
 \foreach \x in {1,2,3}
  \draw (\x,0) -- (\x,3);
 \foreach \y in {0,1,2,3}
  \draw (0,\y) -- (4,\y);
 \foreach \y in {0,1,2}
 {
  \draw[fill=black!10] (0,\y) -- (1,\y) -- (0,\y+1) -- cycle;
  \draw[fill=black!10] (4,\y) -- (3,\y) -- (4,\y+1) -- cycle;
 }
 \foreach \x in {1.5,2.5}
 \foreach \y in {0,1,2,3}
 {
  \draw[thick,red] (\x-0.25,\y-0.25) -- (\x+0.25,\y+0.25);
  \draw[thick,red] (\x+0.25,\y-0.25) -- (\x-0.25,\y+0.25);
 }
 \foreach \x in {1,2,3}
 \foreach \y in {0.5,1.5,2.5}
 {
  \draw[thick,red] (\x-0.25,\y-0.25) -- (\x+0.25,\y+0.25);
  \draw[thick,red] (\x+0.25,\y-0.25) -- (\x-0.25,\y+0.25);
 }
 \foreach \y in {0,1,2}
 {
  \draw[thick, red, rounded corners=0.2cm] (-0.25,\y-0.25) -- (-0.6,\y+0.15) -- (0.65,\y+0.15) -- (0.75,\y+0.25);
  \draw[thick, red, rounded corners=0.2cm] (4.25,\y-0.25) -- (4.6,\y+0.15) -- (3.35,\y+0.15) -- (3.25,\y+0.25);
  \draw[thick, red, rounded corners=0.2cm] (0.75,\y-0.25) -- (-0.25,\y+0.75);
  \draw[thick, red, rounded corners=0.2cm] (3.25,\y-0.25) -- (4.25,\y+0.75);
 }
 \draw[thick, red] (0.35,-0.25) -- (0.35,3.1);
 \draw[thick, red] (3.65,-0.25) -- (3.65,3.1);
 \draw[thick, red] (0.75,2.75) -- (0.4,3.1);
 \draw[thick, red] (3.25,2.75) -- (3.6,3.1);
 \draw[thick, red, rounded corners=0.2cm] (-0.25,2.75) -- (-0.6,3.15) -- (0.3,3.15);
 \draw[thick, red, rounded corners=0.2cm] (4.25,2.75) -- (4.6,3.15) -- (3.7,3.15); 
 \node at (-0.25,-0.5) {\tiny $1$};
 \node at (0.25,-0.5) {\tiny $2$};
 \node at (0.75,-0.5) {\tiny $3$};
 \node at (1.25,-0.5) {\tiny $4$};
 \node at (1.75,-0.5) {\tiny $5$};
 \node at (3,-0.5) {\tiny $\cdots$};
 \node at (4.25,-0.5) {\tiny $2N$};
 \foreach \x in {0,1,2,3,4}
 \foreach \y in {0,1,2,3}
 {
  \fill (\x,\y) circle (0.1cm);
 }
\end{tikzpicture}
&\qquad
\begin{tikzpicture}[scale=0.95]
 \foreach \x in {1,2,3}
  \draw (\x,0) -- (\x,3);
 \foreach \y in {0,1,2,3}
  \draw (0,\y) -- (4,\y);
 \foreach \y in {0,1,2}
 {
  \draw[fill=black!10] (0,\y) -- (1,\y) -- (1,\y+1) -- (0,\y+1) -- cycle;
  \draw[fill=black!10] (4,\y) -- (3,\y) -- (3,\y+1) -- (4,\y+1) -- cycle;
 }
\draw[thick, red, rounded corners=0.2cm] (0.4,-0.25) -- (0.25,0) -- (0.75,1) -- (0.25,2) -- (0.75,3) -- (0.6,3.25);
\draw[thick, red, rounded corners=0.2cm] (0.6,-0.25) -- (0.75,0) -- (0.25,1) -- (0.75,2) -- (0.25,3) -- (0.4,3.25);
\draw[thick, red, rounded corners=0.2cm] (3.6,-0.25) -- (3.75,0) -- (3.25,1) -- (3.75,2) -- (3.25,3) -- (3.4,3.25);
\draw[thick, red, rounded corners=0.2cm] (3.4,-0.25) -- (3.25,0) -- (3.75,1) -- (3.25,2) -- (3.75,3) -- (3.6,3.25);
\draw[thick, red, rounded corners=0.2cm] (-0.25,-0.25) -- (-0.5,0) -- (0,0.25) -- (1,0.75) -- (1.5,1) -- (2.5,2) -- (3,2.25) -- (4,2.75) -- (4.5,3) -- (4.25,3.25);
\draw[thick, red, rounded corners=0.2cm] (4.25,-0.25) -- (4.5,0) -- (4,0.25) -- (3,0.75) -- (2.5,1) -- (1.5,2) -- (1,2.25) -- (0,2.75) -- (-0.5,3) -- (-0.25,3.25);
\draw[thick, red, rounded corners=0.2cm] (1.75,-0.25) -- (1.5,0) -- (1,0.25) -- (0,0.75) -- (-0.5,1) -- (0,1.25) -- (1,1.75) -- (1.5,2) -- (2.5,3) -- (2.75,3.25);
\draw[thick, red, rounded corners=0.2cm] (2.25,-0.25) -- (2.5,0) -- (3,0.25) -- (4,0.75) -- (4.5,1) -- (4,1.25) -- (3,1.75) -- (2.5,2) -- (1.5,3) -- (1.25,3.25);
\draw[thick, red, rounded corners=0.2cm] (1.25,-0.25) -- (1.5,0) -- (2.5,1) -- (3,1.25) -- (4,1.75) -- (4.5,2) -- (4,2.25) -- (3,2.75) -- (2.5,3) -- (2.25,3.25);
\draw[thick, red, rounded corners=0.2cm] (2.75,-0.25) -- (2.5,0) -- (1.5,1) -- (1,1.25) -- (0,1.75) -- (-0.5,2) -- (0,2.25) -- (1,2.75) -- (1.5,3) -- (1.75,3.25);
 \node at (-0.25,-0.5) {\tiny $1$};
 \node at (0.25,-0.5) {\tiny $2$};
 \node at (0.75,-0.5) {\tiny $3$};
 \node at (1.25,-0.5) {\tiny $4$};
 \node at (1.75,-0.5) {\tiny $5$};
 \node at (3,-0.5) {\tiny $\cdots$};
 \node at (4.25,-0.5) {\tiny $2N$};
 \foreach \x in {0,1,2,3,4}
 \foreach \y in {0,1,2,3}
 {
  \fill (\x,\y) circle (0.1cm);
 }
\end{tikzpicture}
\nonumber
\end{align}
Let us explicitly write the boundary interactions in the case $S=3$. We label the vertices of a boundary triangle by $i,j,k$, and the corresponding Potts spins by $\sigma_i,\sigma_j,\sigma_k\in\{1,2,\cdots, Q\}$, whose interactions are given by the local Hamiltonian
\begin{align}
-\mathcal{ H} = K_1 \delta_{\sigma_i,\sigma_j} + K_2 (\delta_{\sigma_i,\sigma_k} + \delta_{\sigma_j,\sigma_k}) + K_3 \delta_{\sigma_i,\sigma_j,\sigma_k} \,. \qquad
\begin{tikzpicture}[baseline = (base)]
\coordinate (base) at (0, .5);
 \draw[fill=black!10] (0.0,0.0) -- (1.0,0.0) -- (0,1) -- cycle;
 \draw (0.0,0.0) -- (1.0,0.0) -- (0,1) -- cycle;
 \draw [fill] (0,0) circle (.4ex);
 \draw [fill] (1.0,0.0) circle (.4ex);
 \draw [fill] (0, 1) circle (.4ex);
 \node [left] at (0, 0) {$i$};
 \node [left] at (0, 1) {$j$};
 \node [right] at (1, 0) {$k$};
\end{tikzpicture}
\end{align}
The corresponding Boltzmann weights
\begin{equation}
 \mathrm{ e}^{-\mathcal{ H}} = c_0 + c_1 \delta_{\sigma_i,\sigma_j} + c_2 (\delta_{\sigma_i,\sigma_k} + \delta_{\sigma_j,\sigma_k}) + c_3 \delta_{\sigma_i,\sigma_j,\sigma_k}
\end{equation}
are related to the coupling constants by
\begin{subequations}
\begin{align}
 1 &= c_0 \,, \\
 \mathrm{ e}^{K_1} &= 1 + c_1 \,, \\
 \mathrm{ e}^{K_2} &= 1 + c_2 \,, \\
 \mathrm{ e}^{K_3 + 2 K_2 + K_1} &= 1 + c_1 + 2 c_2 + c_3 \,.
\end{align}
\end{subequations}
Let us represent the contributions in terms of FK clusters. (We now draw the triangles as equilateral instead of isosceles.)
\begin{align}
\begin{tikzpicture}[rotate=30]
 \draw (0.0,0.0) -- (1.0,0.0) -- (0.5,0.866) -- cycle;
 \draw [fill] (0,0) circle (.5ex);
 \draw [fill] (1.0,0.0) circle (.5ex);
 \draw [fill] (0.5,0.866) circle (.5ex);
 \draw (0.2,-0.6) node {$c_0$};
\end{tikzpicture} \quad
\begin{tikzpicture}[rotate=30]
 \draw (0.0,0.0) -- (1.0,0.0) -- (0.5,0.866) -- cycle;
 \draw[blue,ultra thick] (0,0) -- (0.5,0.289) -- (0.5,0.866);
 \draw [fill] (0,0) circle (.5ex);
 \draw [fill] (1.0,0.0) circle (.5ex);
 \draw [fill] (0.5,0.866) circle (.5ex);
 \draw (0.2,-0.6) node {$c_1$};
\end{tikzpicture} \quad
\begin{tikzpicture}[rotate=30]
 \draw (0.0,0.0) -- (1.0,0.0) -- (0.5,0.866) -- cycle;
 \draw[blue,ultra thick] (0,0) -- (0.5,0.289) -- (1.0,0.0);
 \draw [fill] (0,0) circle (.5ex);
 \draw [fill] (1.0,0.0) circle (.5ex);
 \draw [fill] (0.5,0.866) circle (.5ex);
 \draw (0.2,-0.6) node {$c_2$};
\end{tikzpicture} \quad
\begin{tikzpicture}[rotate=30]
 \draw (0.0,0.0) -- (1.0,0.0) -- (0.5,0.866) -- cycle;
 \draw[blue,ultra thick] (0,0) -- (0.5,0.289) -- (0.5,0.866);
 \draw[blue,ultra thick] (0.5,0.289) -- (1.0,0.0);
 \draw [fill] (0,0) circle (.5ex);
 \draw [fill] (1.0,0.0) circle (.5ex);
 \draw [fill] (0.5,0.866) circle (.5ex);
 \draw (0.2,-0.6) node {$c_3$};
\end{tikzpicture}
\end{align}
or, equivalently, in terms of loops bouncing off the clusters as
\begin{align}
\begin{tikzpicture}[rotate=30]
 \draw (0.0,0.0) -- (1.0,0.0) -- (0.5,0.866) -- cycle;
 \draw [fill] (0,0) circle (.5ex);
 \draw [fill] (1.0,0.0) circle (.5ex);
 \draw [fill] (0.5,0.866) circle (.5ex);
 \draw [thick,red,rounded corners=2mm] (-0.166,0.289) -- (0.5,0.289) -- (0.167,-0.289);
 \draw [thick,red,rounded corners=2mm] (0.167,0.866) -- (0.5,0.289) -- (0.833,0.866);
 \draw [thick,red,rounded corners=2mm] (1.166,0.289) -- (0.5,0.289) -- (0.833,-0.289);
 \draw (0.2,-0.6) node {$a_0$};
\end{tikzpicture} \quad
\begin{tikzpicture}[rotate=30]
 \draw (0.0,0.0) -- (1.0,0.0) -- (0.5,0.866) -- cycle;
 \draw [fill] (0,0) circle (.5ex);
 \draw [fill] (1.0,0.0) circle (.5ex);
 \draw [fill] (0.5,0.866) circle (.5ex);
 \draw [thick,red,rounded corners=2mm] (-0.166,0.289) -- (0.5,0.289) -- (0.167,0.866);
 \draw [thick,red,rounded corners=2mm] (0.167,-0.289) -- (0.833,0.866);
 \draw [thick,red,rounded corners=2mm] (0.833,-0.289) -- (0.5,0.289) -- (1.166,0.289);
 \draw (0.2,-0.6) node {$a_1$};
\end{tikzpicture} \quad
\begin{tikzpicture}[rotate=30]
 \draw (0.0,0.0) -- (1.0,0.0) -- (0.5,0.866) -- cycle;
 \draw [fill] (0,0) circle (.5ex);
 \draw [fill] (1.0,0.0) circle (.5ex);
 \draw [fill] (0.5,0.866) circle (.5ex);
 \draw [thick,red,rounded corners=2mm] (1.166,0.289) -- (0.5,0.289) -- (0.833,0.866);
 \draw [thick,red,rounded corners=2mm] (0.167,0.866) -- (0.833,-0.289);
 \draw [thick,red,rounded corners=2mm] (0.167,-0.289) -- (0.5,0.289) -- (-0.166,0.289);
 \draw (0.2,-0.6) node {$a_2$};
\end{tikzpicture} \quad
\begin{tikzpicture}[rotate=30]
 \draw (0.0,0.0) -- (1.0,0.0) -- (0.5,0.866) -- cycle;
 \draw [fill] (0,0) circle (.5ex);
 \draw [fill] (1.0,0.0) circle (.5ex);
 \draw [fill] (0.5,0.866) circle (.5ex);
 \draw [thick,red,rounded corners=2mm] (0.167,-0.289) -- (0.5,0.289) -- (0.833,-0.289);
 \draw [thick,red,rounded corners=2mm] (0.833,0.866) -- (0.5,0.289) -- (1.166,0.289);
 \draw [thick,red,rounded corners=2mm] (-0.166,0.289) -- (0.5,0.289) -- (0.167,0.866);
 \draw (0.2,-0.6) node {$a_3$};
\end{tikzpicture}
\end{align}
The two sets of weights are related by the Euler relation:
\begin{align}
 (a_0,a_1,a_2,a_3)=\left(c_0,n^{-1}c_1,n^{-1}c_2,n^{-2}c_3\right)\ .
\end{align}
Matching this with the algebraic expression for the projector
\begin{equation}
 P^\mathrm{ left}_3 = 1 - \tfrac{n}{n^2-1} (e_1+ e_{2}) + \tfrac{1}{n^2-1} ( e_1 e_{2} + e_{2} e_1) \,,
\end{equation}
we find the weights
\begin{align}
 (a_0,a_1,a_2,a_3)= \left(-\tfrac{n}{n^2-1},1,\tfrac{1}{n^2-1},-\tfrac{n}{n^2-1}\right)\ .
\end{align}
Renormalising such that $c_0=1$, this corresponds to
\begin{align}
 (c_0,c_1,c_2,c_3)= (1, 1-Q, -1, Q)\ .
\end{align}
The coupling constants are therefore
\begin{align}
 \mathrm{ e}^{K_1} = 2-Q \quad , \quad
 \mathrm{ e}^{K_2} = 0 \,,
\end{align}
whereas $\mathrm{ e}^{K_3}$ does not matter, since $e^{K_2}=0$ prevents the three spins from ever being equal.

% Notice that for $Q=2$ all three spins in the boundary triangle must be pairwise different, which is impossible, so
% the $S=3$ boundary condition is singular in the limit $Q \to 2$.  Note also that $\mathrm{ e}^{K_1}<0$ for $Q>2$. In this case, the corresponding coupling in the Potts model must be taken to have an imaginary part equal to $\pi$.

\subsubsection*{Agreement with the CFT}

From a transfer matrix $T$, we deduce the critical limit of the partition function on an annulus of modulus $\tau$,
\begin{align}
 Z=\lim_\text{critical} \operatorname{Tr} T^M \qquad \text{with} \qquad \tau = \lim_\text{critical}i\frac{M}{2N} \ .
\end{align}
In practice, we test the agreement with the CFT by diagonalising the transfer matrix, whose eigenvalues should correspond to the conformal dimensions that appear in the CFT's boundary spectrum.
The boundary channel expansion of the CFT partition function $Z_{S,S}$ \eqref{zss} predicts eigenvalues of the type $\Delta_{(r,s)}+L$ with $r\in\mathbb{N}^*$ and $s=1,3,\cdots, 2S-1$ and $L\in\mathbb{N}$.

We have numerically computed the $40$ leading eigenvalues for each of the boundary conditions $S=2,3,4$
with $r=0,1,2,3$ and sizes up to $N=14$. The diagonalisation of $T_{S,S}$ is done in the basis of standard modules
(link patterns) with $2r$ defect lines. In general the spectra appear to be real and simple, provided that $Q$
is taken sufficiently large. For instance, with $r=0$ and $Q=2.5$, we found that $T_{3,3}$ has a real spectrum.
% Notice that this is larger than the limit $Q \to 2$ for which the boundary interaction was found to be singular.
Resolving carefully all level crossings of the finite-size precursors of scaling levels upon varying $N$, and then
extrapolating the finite-size estimates of the exponents to $N \to \infty$, we were able to
identify the first $30$ or $35$ scaling levels.
Perfect agreement with the CFT predictions $Z_{S,S}$ from \eqref{zss} was found in all cases: the scaling levels agree with the CFT's conformal dimensions, and come with the correct degeneracies.

\subsection{Classification of known boundary conditions}\label{sec:ckbc}

For any known boundary condition in critical loop models, we wish to determine if it is reflecting, and if yes the value of its boundary parameter. It turns out that all known reflecting boundaries are discrete with parameter $S\in\{1,2\}$. We now summarise our findings in a table that indicates the specific model where each boundary condition was constructed. Loop models include the $O(n)$, Potts and $PSU(n)$ models; there are also generic constructions that work in all these models, in which case we do not specify a model:
\begin{align}
 \begin{tabular}{|c|c|c|c|}
  \hline
  Model & Name & Type & References
  \\
  \hline \hline
  & free & $S=1$ & \cite{BauerSaleur89,Cardy06,ars22}
  \\
  \hline
  $3$-state Potts & new & $S=2$ & \cite{Ian_Affleck_1998}
  \\
  \hline
  dilute $O(n)$ with $K_b<K_s$ & free (ordinary) & $S=1$ &  \cite{Cardy84, fs94, by94, by95b}
  \\
  \hline
  dilute $O(n)$ with $K_b=K_s$ & special & $S=2$ &  \cite{Cardy84, fs94, by94, by95b}
  \\
  \hline
   & JS & sticky & \cite{js06,js07,djs08}
   \\
   \hline
   dilute $O(n)$ with $K_b = K_s(n_1)$ & JS-special & sticky & \cite{djs09a,djs09}
   \\
   \hline
   Potts
   & fixed & sticky & \cite{Ian_Affleck_1998}
   \\
   \hline
   Potts
  & wired & $S=1$ & \cite{Ian_Affleck_1998}
  \\
  \hline
 \end{tabular}
\end{align}

\subsubsection*{Dilute $O(n)$ model}

For $1 < \beta^2 \le 2$, the critical $O(n)$ model is in a dilute phase. The model depends not only on the loop weight $n$, but also on the monomer weight $K$, which has the critical value
$K_c=\frac{1}{\sqrt{2+\sqrt{2-n}}}$ on a hexagonal lattice \cite{nie82}.
In the presence of a boundary, we can give a different weight $K_b$ to boundary monomers, and the critical limit depends on $K_b$.
There is a value $K_s=(2-n)^{-\frac14}>K_c$ \cite{by95} such that for $K_b<K_s$ we obtain the free boundary condition (also called ordinary), and for $K_b=K_s$ we obtain a different boundary condition called special.

\subsubsection*{Sticky boundaries}

If a boundary is sticky, it is possible to change the weights of loops that are attached to the boundary. Instead of the loop weight $n$, such loops can have an arbitrary weight $n_1\in\mathbb{C}$. This gives rise to the so-called JS boundary conditions.
There is a variant for the dilute $O(n)$ model, called JS-special in the table (and anisotropic special in the original references \cite{djs09a,djs09}), for which the special monomer weight $K_s$ depends on both $n$ and $n_1$.

Either type of JS boundary condition is sticky in the sense of the diagrams in the central panel of (1.1), not because a piece of loop approaching the boundary can turn into an end point of an arc (middle part of the figure), but rather because the weight of the whole loop depends on whether it does or does not touch the boundary (left and right parts of the figure).

We can also define boundaries where individual arcs can end \cite{js06}, or where any number of arcs can end at the same point.

\subsubsection*{Global symmetry}

The $O(n)$, Potts and $PSU(n)$ loop models have global symmetries that include (but are much larger than) the groups $O(n)$, $S_Q$ and $PSU(n)$ \cite{rs07}. These global symmetries commute with the Jones--Wenzl projector that appears in our transfer matrix \eqref{TS1S2}, and are left unbroken by modifying the weights of boundary monomers as was done in the dilute $O(n)$ model.
On the other hand, in the $O(n)$ model, giving weight $n_1$ to loops that are attached to the boundary can be interpreted as breaking the symmetry from $O(n)$ down to $O(n_1)\times O(n-n_1)$. Similarly, allowing arcs to end on the boundary can be interpreted as adding an external field there, which breaks the global symmetry down to $O(n-1)$.

This suggests that the global symmetry is preserved by reflecting boundaries, and broken by sticky boundaries. This suggestion is supported by the structure of the CFT spectrum, where diagonal fields are global symmetry singlets, whereas non-diagonal fields transform in non-trivial representations  of the global symmetry \cite{jrs22, rjrs24}.

In the Potts model, fixed boundary conditions are defined by setting boundary spins to a fixed value. Of course, this breaks global symmetry, which relates all possible values of the spin. And indeed, fixed boundary conditions are found to be sticky. (Except if $Q=1$, when the spin has only one possible value.) It is, however, possible to restore the symmetry by summing over all values of
the fixed boundary spin: this is called a wired boundary condition, because it is equivalent to contracting the whole boundary to a single spin. Fixed and wired boundary conditions were studied in detail in the case of the $3$-state Potts model \cite{Ian_Affleck_1998}. It was found that free and wired boundary conditions are dual to one another, and therefore equivalent from the CFT's point of view. And indeed, we find that free and wired boundaries are both reflecting with the same parameter $S=1$.

\subsubsection*{Arcs at the boundary}

To this point we have discussed lattice boundary conditions involving only loops. Boundary conditions where arcs can end on the boundary
--- and which, from the field-theoretic point of view, correspond to Dirichlet boundary conditions for the $O(n)$ vector spin --- have been less studied in the literature
(but see \cite{djs09} for an exception). More detail on such boundary conditions will appear elsewhere.

\section{Conclusion and outlook}\label{sec:con}

\subsubsection*{Characterising reflecting boundaries}

From our definition and study of boundary conditions in critical loop models, there emerge several equivalent criteria for a boundary to be reflecting:
\begin{enumerate}
\item Loops are never attached to the boundary, and arcs cannot end on it.
 \item The disc one-point functions of non-diagonal fields vanish.
 \item In the bulk channel, the annulus partition function has no contribution from non-diagonal fields.
 \item The boundary condition preserves the global symmetry of the loop model.
\end{enumerate}
We have determined the space of reflecting boundary conditions: this space is parametrised by a complex number $\sigma\in\mathbb{C}$, and includes a subset parametrised by $S\in\mathbb{N}^*$, corresponding to discrete boundaries. In this space, only the two cases $S=1$ and $S=2$ were previously studied. Nevertheless, we could provide analytic expressions for correlation functions in the presence of arbitrary reflecting boundaries, and a lattice construction of all discrete reflecting boundaries.

\subsubsection*{Combinatorial description of correlation functions}

In bulk critical loop models, correlation functions are parametrised by combinatorial maps \cite{gjnrs23}. We expect that this remains true in the presence of boundaries. In the case of bulk 2-point functions on the disc with a discrete reflecting boundary, our results agree with this expectation. For example, in the case of the 2-point function $\left<V_{(\frac32, 0)}V_{(\frac12,0)}\right>$, the fields that appear have 3 legs and 1 leg respectively, leading to a unique combinatorial map on the disc:
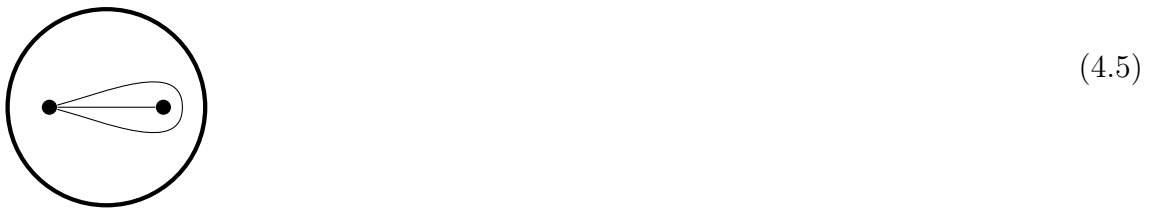
\begin{align}
\begin{tikzpicture}[baseline=(base), scale = .5]
    \coordinate (base) at (0, .8);
  \node[fill, circle, minimum size = 2mm, inner sep = 0] at (0, 0) (a) {};
  \node[fill, circle, minimum size = 2mm, inner sep = 0] at (3, 0) (c) {};
    \draw (a) -- (c);
    \draw (a) to [out = -15, in = -90] (3.5, 0) to [out = 90, in = 15] (a);
    \draw (1.5, 0) ellipse (2.6cm and 2.6cm);
  \end{tikzpicture}
\end{align}
The uniqueness of this map corresponds to the uniqueness of the solution of crossing symmetry for this 2-point function. At the combinatorial level, the problem can be mapped to the sphere by replacing the disc's boundary with a legless vertex. A disc with 2 vertices (whose total number of legs is even) gives rise to a unique map, just like a sphere with 3 vertices. More generally, we have the equivalence
\begin{align}
 \left<\prod_{i=1}^N V_{(r_i,s_i)}\right>^\text{disc with reflecting boundary} \underset{\text{combinatorially}}{\simeq} \left<V_P\prod_{i=1}^N V_{(r_i,s_i)}\right>^\text{sphere}\ .
\end{align}
This allows us to predict the dimension of the space of solutions of crossing symmetry, for any correlation function of bulk fields on the disc.

\subsubsection*{Further bootstrap challenges}

Having determined disc 1-point functions for reflecting boundaries, we think we understand the space of such boundaries, with its coordinate $\sigma\in\mathbb{C}$. We can compute disc 2-point functions of bulk fields in the bulk channel, and we could compute disc $N$-point functions if we knew the bulk OPE. In order to solve critical loop models on the disc, we still need to understand boundary fields and the associated structure constants: the bulk-boundary 2-point structure constant, and the boundary 3-point structure constant. This problem can be addressed using both the analytic and numerical bootstrap methods.

Then there is the challenge of sticky boundaries. This is more difficult than the reflecting case, because we do not have a shift equation that would constrain how the disc 1-point function $\left<V_{(r,0)}\right>$ depends on $r$. Moreover, since legs can end at the boundary, the combinatorics predict large numbers of solutions of crossing symmetry.

We may look for hints of the existence and properties of sticky boundaries in the annulus partition functions of \cite{djs08}. These partition functions depend on several continuous parameters: the weights of various types of loops, depending on their topology and on whether or not they are attached to the boundaries. For generic values of the parameters, in the bulk-channel decomposition of the partition function, there are contributions from non-diagonal fields $V_{(r,0)}$ and also from degenerate fields including $V^d_{\langle 1,2\rangle}$. If we could factorise the annulus partition function into disc 1-point functions, this would contradict Eq. \eqref{mv}. However, in the presence of non-diagonal fields, we know that bulk OPEs are problematic, which suggests that we cannot deduce disc 1-point functions from annulus partition functions. Nevertheless, since we do know partition functions for an annulus with reflecting boundaries, we can conclude that the boundaries of \cite{djs08} are generically sticky.

\setcounter{secnumdepth}{0}

\section{Acknowledgements}

\begin{itemize}
 \item We are grateful to Paul Roux, Xin Sun and Gérard Watts for stimulating discussions.
 \item We thank Gérard Watts and Paul Roux for carefully reading a draft version of the manuscript and providing useful comments.
 \item We are grateful to the SciPost editor and referees for their work on this article.
\item RN thanks IPhT for its hospitality while part of this work was completed.
\item MD is supported by the French Agence Nationale de la Recherche (ANR) under grant ANR-21-CE40-0003 (project CONFICA).
\end{itemize}

\bibliographystyle{inputs/morder8}
\bibliography{inputs/992}

\end{document}